\documentclass[final,onefignum,onetabnum,dvipsnames]{siamart171218}

\usepackage[utf8]{inputenc}
\usepackage{amssymb}
\usepackage{siunitx}
\usepackage{ushort} % short underline
\usepackage{multirow, makecell}
\usepackage{subcaption}
\usepackage{booktabs}
\usepackage{xcolor}

\usepackage{pgf,tikz,pgfplots}
\pgfplotsset{compat=1.15}
\usetikzlibrary{graphs,patterns,arrows,decorations}
\usetikzlibrary{decorations.pathreplacing}
\usetikzlibrary{decorations.pathmorphing}
\tikzstyle{stepnum}=[shade,shading=ball,ball color=gray!25!white,circle,draw,minimum size=1.5em,inner sep=1]
\tikzstyle{stagevalue}=[fill,circle,minimum size=5pt,inner sep=1,draw]
\tikzstyle{store} = [pin edge={-to,blue}]
\tikzstyle{restore} = [pin edge={to-,red}]
\tikzstyle{restorefromdisk} = [pin edge={to-,thick,black},pin distance=0.3cm]
\tikzstyle{restorekeepinstack} = [pin edge={to-,red,dashed}]
\tikzstyle{edgefrom} = [<-,thick]
\tikzstyle{edgebend} = [->,thick,>=stealth',bend right=60]
\usepackage{tikzscale}
\usepackage{fontawesome} % needed by tikz
\usepackage{filecontents}

\usepackage{listings}
\usepackage{bm}

\newcommand{\bu}{\bm{u}}
\newcommand{\bU}{\bm{U}}
\newcommand{\bUref}{\bm{U}_r}
\newcommand{\bp}{\bm{p}}

\newcommand{\cS}{\bm{\mathcal{S}}}

\newcommand{\bw}{\bm{w}}
\newcommand{\bv}{\bm{v}}

\newcommand{\R}{\mathbb{R}}

\newcommand{\boldeta}{\bm{\mathcal{\eta}}}

\newcommand{\M}{\bm{\mathcal{M}}}
\newcommand{\N}{\bm{\mathcal{N}}}

\newcommand{\blambda}{\bm{\lambda}}
\newcommand{\bLambda}{\bm{\Lambda}}
\newcommand{\bGamma}{\bm{\Gamma}}
\newcommand{\bmu}{\bm{\mu}}
\newcommand{\f}{\bm{f}}

\newcommand{\fp}{\f_{\bp}}
\newcommand{\fu}{\f_{\bu}}
\newcommand{\fuu}{\f_{\bu\bu}}
\newcommand{\fup}{\f_{\bu\bp}}
\newcommand{\fpu}{\f_{\bp\bu}}
\newcommand{\fpp}{\f_{\bp\bp}}

\newcommand{\F}{\bm{F}}
\newcommand{\G}{\bm{G}}

\newcommand{\Fu}{\F_{\bu}}

\newcommand{\vect}[1]{\left[ \begin{array}{c} #1 \end{array} \right]}
\newcommand{\DDPSI}[3]{\frac{\partial}{\partial #1}\left(\frac{\partial \psi}{\partial #2}#3\right)^T}
\newcommand{\bsigma}{\bm{\sigma}}

\newif\iflongversion
\longversionfalse

\newsiamremark{remark}{Remark}
\newsiamremark{hypothesis}{Hypothesis}
\crefname{hypothesis}{Hypothesis}{Hypotheses}
\newsiamthm{claim}{Claim}

\headers{PETSc TSAdjoint}{H. Zhang, E. M. Constantinescu and B. F. Smith}

\title{PETSc TSAdjoint: a discrete adjoint ODE solver for first-order and second-order sensitivity analysis \thanks{Submitted to the editors DATE.}
\funding{This material is based upon work supported by the U.S. Department of
Energy, Office of Science, Office of Advanced Scientific Computing Research,
Scientific Discovery through Advanced Computing (SciDAC) program through the
FASTMath Institute under contract DE-AC02-06CH11357 at Argonne National
Laboratory.}}

\author{
  Hong Zhang\thanks{Argonne National Laboratory, Lemont, IL
  (\email{hongzhang@anl.gov}).}
  \and
  Emil M. Constantinescu\thanks{Argonne National Laboratory, Lemont, IL
  (\email{emconsta@anl.gov}).}
  \and
  Barry F. Smith\thanks{Argonne National Laboratory, Lemont, IL
  (\email{bsmith@petsc.dev}).}
}

\begin{document}
\maketitle
\begin{abstract}
We present a new software system \texttt{PETSc} \texttt{TSAdjoint} for
first-order and second-order adjoint sensitivity analysis of time-dependent
nonlinear differential equations. The derivative calculation in \texttt{PETSc}
\texttt{TSAdjoint} is essentially a high-level algorithmic differentiation
process. The adjoint models are derived by differentiating the timestepping
algorithms and implementing them based on the parallel infrastructure in
\texttt{PETSc}. Full differentiation of the library code, including MPI routines,
is avoided, and users do not need to derive their own adjoint models for
their specific applications. \texttt{PETSc} \texttt{TSAdjoint} can compute the
first-order derivative, that is, the gradient of a scalar functional, and the
Hessian-vector product, which carries second-order derivative information, while
requiring minimal input (a few callbacks) from the users. The adjoint model employs optimal checkpointing
schemes in a manner that is transparent to
users. Usability, efficiency, and scalability are demonstrated through examples
from a variety of applications.
\end{abstract}

% REQUIRED

% REQUIRED

\section{Introduction}

Adjoint methods have been used extensively in computational modeling and
optimization, playing a key role in neural networks, sensitivity analysis,
goal-oriented error estimation, data assimilation, and optimal control. They are
efficient algorithmic differentiation (AD) approaches for computing the
derivatives of an objective function of the solution of an ordinary differential
equation (ODE) or differential-algebraic equation (DAE) with respect to
parameters of interest, with a cost independent of the number of parameters.
Deriving the adjoint model is trivial for linear models but can be difficult
for nonlinear models \cite{Farrell2013}, especially time-dependent problems.

Many tools have been developed to derive and implement adjoint models automatically. These automatic tools take as input a forward model that users implement
in languages such as C \cite{Bischof1997,Walther2012}, C++ \cite{Bartlett2006},
Fortran \cite{Giering1998}, Python \cite{Paszke2017}, and Julia \cite{Lubin2015};
and they produce as output the associated discrete adjoint model in a line-by-line
fashion, through source-to-source transformations, operator overloading, or a
combination of both. While this black-box approach gives the highest degree of
automation and requires the least knowledge of the mathematical models, it
suffers from many low-level implementation-specific difficulties including
memory allocation, management of pointers, input/output, and parallel
communication (e.g., MPI and OpenMP). Such black-box tools also often produce far
from optimally efficient code.

Traditional AD treats a model as a sequence of primitive instructions (e.g.,
addition, multiplication, logarithm) and calculates the derivatives based on
the chain rule using the derivatives of these primitive instructions, which are
easily obtainable. In order to overcome the difficulties of these low-level approaches,
 high-level AD libraries such as \texttt{dolfin-adjoint}
\cite{Farrell2013} and \texttt{FATODE} \cite{Zhang2014} recently have been developed to
operate at high abstraction levels.

\begin{figure}
  \centering
  \includegraphics[width=0.86\linewidth]{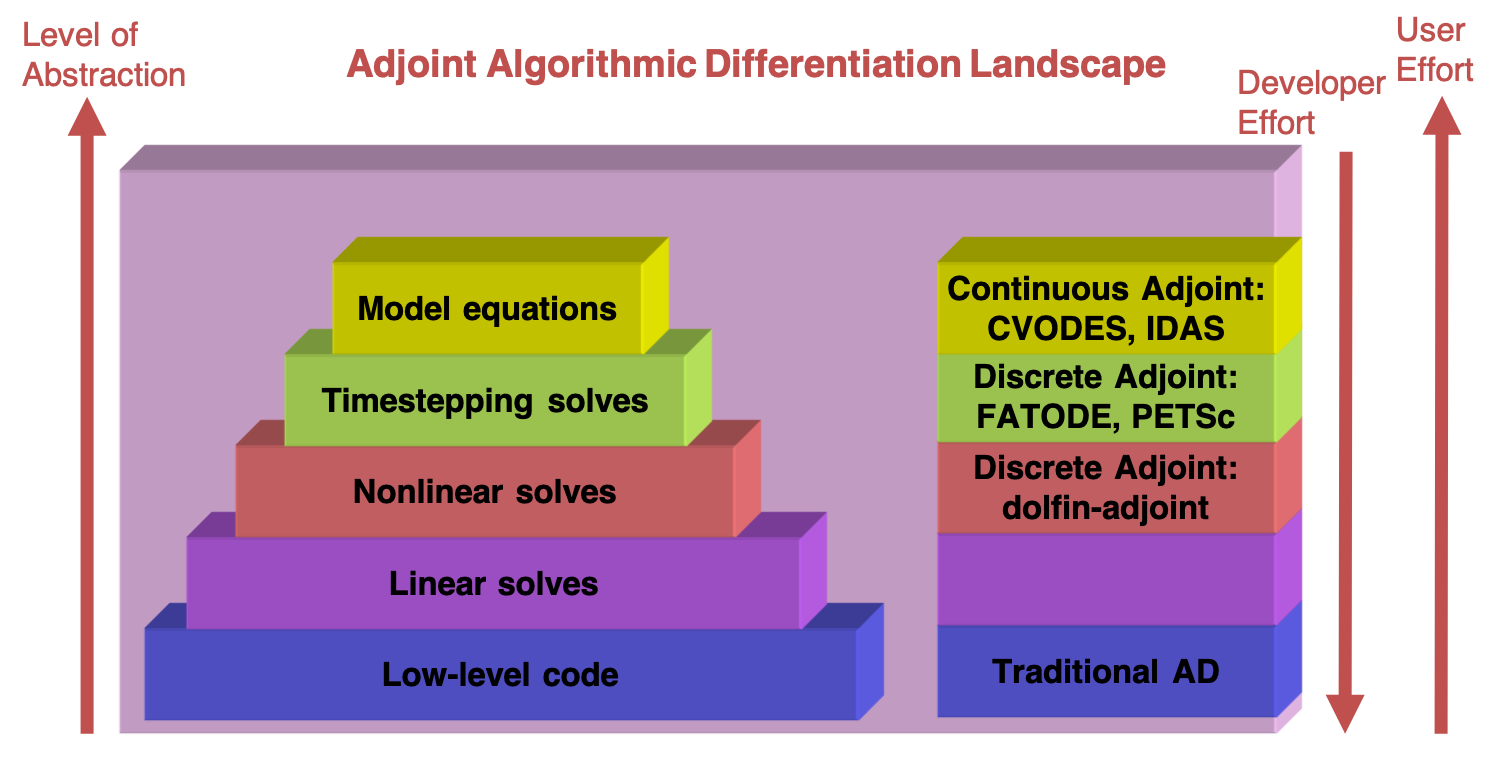}
  \caption{Landscape of adjoint algorithmic differetiation software.}
  \label{fig:ad_abstraction}
\end{figure}

The landscape of popular existing AD software is depicted in Figure
\ref{fig:ad_abstraction}. While these software packages are developed based on
the same theory, they differ significantly in usage and require varying levels
of effort from developers and users. \texttt{Dolfin-adjoint}
\cite{Farrell2013} considers a model as a sequence of nonlinear equation solves
in the form $A(u)u = b(u)$, where $u$ is the vector of all prognostic variables,
$b(u)$ is the source term, and $A(u)$ is the entire discretization matrix. The
derivation of the adjoint model is fully automated in \texttt{dolfin-adjoint} if
the forward model is written in a high-level language that is similar to
mathematical notation. \texttt{Dolfin-adjoint} is used primarily by finite-element systems such as \texttt{FEniCS} \cite{fenics2015} and \texttt{Firedrake}
\cite{Firedrake2016}. \texttt{FATODE} implements an adjoint model by considering
the algorithm of solving time-dependent differential equations as a sequence of
timestepping solves. \texttt{FATODE} provides a built-in implementation of the
adjoint model derived based on the timestepping algorithms for solving ODEs; simulation of time-dependent partial differential
equations (PDEs) is abstracted as a sequence of time steps, and the libraries
differentiate each time step. In contrast, the adjoint solvers \texttt{CVODES}
and \texttt{IDAS} in the \texttt{SUNDIALS} \cite{Hindmarsh2005} package, which
have been used by notable optimization tools such as CasADi
\cite{Andersson2019}, implement an adjoint model that users derive directly from
the model equations. This highest-level approach, also known as the continuous
adjoint approach,  requires users to derive a new set of equations before
discretization (adjoint model of the original continuum or weak form mode). All
the other aforementioned approaches are discrete adjoint approaches since the
adjoint models are derived after discretization. In general, lower-level
abstractions tend to impose more implementation burden on library developers and
provide more automation to users while, at the same time, hiding more
mathematical structures from users. Nevertheless, low-level AD can be mixed with
high-level AD to improve scaling. For example, the internal Jacobian-vector
product in a high-level AD implementation can be effectively computed by using the
traditional reverse-mode AD. This approach has been adopted in many works
\cite{Andersson2019,Rackauckas2018,wallwork2019}.

Another tool that has recently been developed is a Julia library called
DifferentialEquations.jl \cite{Rackauckas2017}. It has support for both
continuous and discrete first-order adjoint sensitivity analysis for ODEs, and
experimental support for second-order sensitivity analysis is claimed.

Adopting a similar approach to that used by \texttt{FATODE}, we have developed
the \texttt{TSAdjoint} component in the Portable, Extensible Toolkit for
Scientific Computation \texttt{PETSc} \cite{Abhyankar2018,Zhang2017}.
\texttt{PETSc} \texttt{TSAdjoint} enables first- and second-order adjoint
sensitivity analysis for nonlinear time-dependent differential equations, which
are the key ingredients of many optimization algorithms. The adjoint models are
derived and implemented for various time integrators in \texttt{PETSc} in a
manner that is agnostic to the spatial discretization engine, thus being
suitable for general-purpose applications. The adjoint models also employ the
parallel infrastructure and the sophisticated linear/nonlinear solvers in
\texttt{PETSc} in the same way as the forward models. Optimal adjoint
checkpointing schemes are implemented and tailored to the needs of the ODE/DAE
solvers. The adjoint control flow is managed automatically by \texttt{PETSc} and
is transparent to users. These features are significant advantages in achieving the
efficiency of adjoint calculation compared with other adjoint codes. In
particular, \texttt{dolfin-adjoint} covers less general applications.
\texttt{FATODE}, \texttt{CVODES}, \texttt{IDAS} and DifferentialEquations.jl
lack support for optimal checkpointing; and to the best of our
knowledge, there are no
reported demonstrations or attempts to use state-of-the-art linear/nonlinear
solvers with these tools, although they are generic to the internal linear
solver. A drawback with \texttt{TSAdjoint} is that the adjoint of each
timestepping algorithm must be implemented by the library developers.

In the next section, we provide the mathematical foundations of sensitivity
analysis for ODE integrators. In Section 3, we explain the software
infrastructure. Section 4 discusses the management of the required checkpointing, and Section 5 explores the use of these algorithms in three examples. Section 6 summarizes our conclusions.

\section{Mathematical Foundation}

This section explains how the sensitivity propagation equations are
derived based on the model abstraction at the timestepping level. Both
first-order and second-order sensitivity analysis approaches are covered. An
example using theta timestepping methods, for which the adjoint model has
moderate complexity, is given to illustrate the details of the derivation. The
mathematical framework can be naturally extended to other timestepping
algorithms, including explicit schemes and even implicit-explicit schemes.

The goal of sensitivity analysis of a dynamical system is to compute the
derivative of a scalar functional with respect to specific system parameters. We consider the dynamical
system in DAE form for notational brevity and without loss of generality:
\begin{equation}
\M \dot{\bu} = \f(t,\bu;\bp)\,,
\label{eq:ode_def}
\end{equation}
where $\M \in \R^{N_d \times N_d}$ is the mass matrix, $\bu \in \R^{N_d}$ is the
system state, and $\bp \in \R^{N_p}$ are the parameters of interest. These forms
typically arise from the semi-discretization of time-dependent PDEs using the method
of lines. The mass matrix may be the identity for typical ODEs or a singular
matrix for DAEs. In this paper, vectors and matrices are denoted by bold letters
and scalars by non-bold letters. The \textit{numerator layout} notation is used
for derivatives; for example, the derivative of a scalar function is a row vector.

Consider time integration as a sequence of operations,
\begin{equation}
\bu_{n+1} = \N (\bu_n), \quad n = 0,\dots,N-1,
\label{eqn:timestepping_operator}
\end{equation} 
where the initial condition is $\bu_0 =\boldeta$ and $\N$ is a timestepping
operator that propagates the solution from $t_n$ to $t_{n+1}$. An example of
$\N$, an implicit timestepping method, is discussed in Section
\ref{sec:Theta:method:example}.
The scalar functional in sensitivity analysis depends on the system states and
is denoted by $\psi(\bu_N)$ if it is a function of the final state or expressed
in integral form
\begin{equation}
  \label{eq:int:form}
\int_{t_0}^{t_F} r (t,\bu;\bp) dt
\end{equation}
if it is a function of the entire trajectory of the system.

In the following two subsections, we briefly explain how the derivatives of a
scalar function $\psi(\bu_N)$ with respect to the initial condition are derived
in the discrete regime. The derivatives with respect to parameters (e.g., model
parameters) can be derived with the same framework by augmenting the parameters
into the initial condition vector. We refer readers to \cite{Zhang2014} for
details.

\subsection{First-order discrete derivatives}
We use the Lagrange multipliers $\blambda_n \in \R^{N_d}, n=0,\dots,N$,
which are column vectors, to account for the constraint from each time step and  
define the Lagrangian
\begin{equation}
\mathcal{L}(\boldeta) = \psi(\bu_N)  - \blambda_0 ^T \left( \bu_0 - \boldeta \right) - \sum_{n=0}^{N-1} \blambda_{n+1} ^T \left( \bu_{n+1} - \N (\bu_n) \right).
\label{eqn:lag}
\end{equation}
We choose the transpose for the convenience of derivation because the
derivative of a row vector with respect to a column vector is not well defined
in matrix calculus.

Taking the total derivative of equation \eqref{eqn:lag} with
respect to the initial condition $\boldeta$ leads to
\begin{equation}
\dfrac{d \mathcal{L} }{d \boldeta} = \blambda_0 ^T   - \left( \frac{d \psi}{d \bu}(\bu_N) -  \blambda_{N} ^T \right)  \frac{d \bu_N}{d \boldeta} - \sum_{n=0}^{N-1}\left( \blambda_{n}^T - \blambda_{n+1} ^T \frac{d \N }{d \bu} (\bu_n) \right)  \frac{d \bu_n}{d \boldeta} .
\label{eq:lag_diff}
\end{equation}
The first-order adjoint equation is defined as
\begin{equation}
  \begin{aligned}
  \blambda_{n} &=  \left(\frac{d \N}{d \bu}(\bu_n)\right)^T \blambda_{n+1}, \quad n= N-1, \dots, 0, \\
  \blambda_{N} &= \left(\frac{d \psi}{d \bu}(\bu_N)\right)^T,
  \end{aligned}
\label{eqn:disadjoint}
\end{equation}
in order to make the last two terms in \eqref{eq:lag_diff} vanish so that the
total derivative can be obtained without computing the forward sensitivities.
Note that in the adjoint model, the sensitivities are calculated by propagating
the derivative information in reverse order.

An alternative approach is to derive the discrete tangent linear model (TLM)
from the discrete forward model. By differentiating directly
\eqref{eqn:timestepping_operator} with respect to the initial condition
$\boldeta$ and defining the sensitivity matrix $\cS_n \in \R^{N_d\times N_d}$ by
\begin{equation}
  \cS_n = \frac{d \bu_n}{d \boldeta},\quad n= 0, \dots, N-1,
\end{equation}
we can obtain the TLM equations 
\begin{equation}
  \cS_{n+1} =  \left( \frac{d \N }{d \bu} (\bu_n) \right) \, \cS_n, \quad n= 0, \dots, N-1,
\end{equation}
which propagates the sensitivity matrix forward in time and can be solved
together with the original model equations \eqref{eq:ode_def}. Similarly, one
can differentiate \eqref{eqn:timestepping_operator} with respect to the
parameters to derive the TLM equations for calculating parameter sensitivities.
These sensitivities can be used to compute the derivative of the scalar
functional through the chain rule so that the TLM method can achieve the same
goal as the adjoint method. However, these two methods may differ significantly
in terms of computational cost. The computational complexity of the adjoint
method is $\mathcal{O}(N_f)$ whereas the complexity of the TLM method is
$\mathcal{O}(N_p)$, where $N_f$ and $N_p$ are the number of functionals and the
number of parameters, respectively. Therefore, the adjoint method is more
efficient than the TLM method when computing derivatives of a scalar functional
with respect to many parameters. The TLM method can be efficient only when there
are few parameters and has limited application compared with the adjoint method.

\subsection{Second-order discrete derivatives}

% \begin{itemize}
% \item forward over forward (FOF)
% \item forward over adjoint (FOA) or adjoint over forward (AOF)
% \item adjoint over adjoint (AOA)
% \end{itemize}
A most computationally efficient approach for calculating second-order
derivatives for a large number of parameters is the forward-over-adjoint method
\cite{Alexe2009}, which requires both the first-order adjoint model and the TLM.
By differentiating the transpose of $\dfrac{d \mathcal{L} }{d \boldeta}$ with
respect to $\boldeta$ for a second time, we obtain
\begin{align}
\dfrac{d}{d \boldeta}\left(\dfrac{d \mathcal{L} }{d \boldeta}\right)^T & = \frac{d \blambda_0}{d \boldeta} - \left(\frac{d \psi}{d \bu}(\bu_N) - \blambda_{N}^T\right) \frac{d^2 \bu_N}{d \boldeta^2} \\
\nonumber
 & -  \left(\frac{d \bu_N}{d \boldeta}\right)^T \left( \frac{d}{d \bu}\left(\frac{d \psi}{d \bu}(\bu_N)\right)^T \frac{d \bu_N}{d \boldeta} - \frac{d \blambda_N}{d \boldeta} \right)  \\
\nonumber
 & - \sum_{n=0}^{N-1} \left(\frac{d \bu_n}{d \boldeta}\right)^T \left( \frac{d \blambda_n}{d \boldeta} - \blambda_{n+1}^T \, \frac{d^2 \N }{d \bu^2} (\bu_n) \, \frac{d \bu_n}{d \boldeta} - \left(\frac{d \N}{d \bu}(\bu_n)\right)^T \frac{d \blambda_n}{d \boldeta} \right) \\
 \nonumber
 & - \sum_{n=0}^{N-1}\left( \blambda_n ^T - \left(\frac{d \N }{d \bu}(\bu_n)\right)^T \blambda_{n+1} \right)  \frac{d^2 \bu_n}{d \boldeta^2}.
\end{align}

By utilizing the first-order adjoint equations \eqref{eqn:disadjoint} and the
second-order adjoint equations
\begin{equation} 
  \begin{aligned}
  \frac{d \blambda_{n}}{d \boldeta} &= \left(\frac{d \N}{d \bu}(\bu_n)\right)^T \frac{d \blambda_{n+1}}{d \boldeta} + 
  \blambda_{n+1}^T \frac{d^2 \N}{d \bu^2}(\bu_n) \frac{\partial \bu_n}{\partial \boldeta}, ~
  n= N-1, \dots, 0, \\
  \frac{d \blambda_{N}}{d \boldeta} &= \frac{d}{d \bu}\left(\frac{d \psi}{d \bu}(\bu_N)\right)^T \frac{\partial \bu_N}{\partial \boldeta},
  \end{aligned}
\label{eqn:2nddisadjoint}
\end{equation}
where $\frac{d \blambda}{d \boldeta}$ carries second-order derivative
information, we obtain the Hessian of the objective function
$
\nabla^2_{\boldeta}\mathcal{L} =  \nabla^2_{\boldeta}\psi(\bu_n) = \frac{d \blambda_0}{d \boldeta}.
$

Equation \eqref{eqn:2nddisadjoint} propagates a matrix, a 
computationally expensive process that is also not storage efficient. Practical
implementations seek to provide the computation of  Hessian-vector products
instead of the full Hessian. To this end, we derive the directional second-order
derivative, which results in a significantly lower complexity. Assume
$\bv \in \R^{N_d}$ is the directional vector that either comes from the optimization
algorithm or is specified by the user. Post-multiplying $v$ on both sides of
\eqref{eqn:2nddisadjoint} gives
\begin{equation} 
  \begin{aligned}
  \frac{d \blambda_{n}}{d \boldeta}\bv &= \left(\frac{d \N}{d \bu}(\bu_n)\right)^T \frac{d \blambda_{n+1}}{d \boldeta} \bv + 
  \blambda_{n+1}^T \frac{d^2 \N}{d \bu^2}(\bu_n) \boxed{\frac{\partial \bu_n}{\partial \boldeta}\bv}, ~
  n= N-1, \dots, 0, \\
  \frac{d \blambda_{N}}{d \boldeta}\bv &= \frac{d}{d \boldeta}\left(\frac{d \psi}{d \bu}(\bu_N)\right)^T \boxed{\frac{\partial \bu_N}{\partial \boldeta}\bv}.
  \end{aligned}
\label{eqn:2nddisadjoint2}
\end{equation} 
The boxed terms in \eqref{eqn:2nddisadjoint2} are the directional
derivatives for the forward sensitivities that can be calculated with a TLM.

These equations can also be derived by differentiating the first-order adjoint
equation \eqref{eqn:disadjoint}. For brevity, we drop $n= N-1, \dots, 0$ in the
adjoint equations in what follows. Readers should keep in mind that the adjoint
equations always go backward in time. Parameters $\bp$ in functions such as $\f$
and $r$ are dropped for the same reason.

\subsection{Augmented system for deriving parametric sensitivity and
incorporating integrals} \label{sec:augmented_sys} 

To obtain the parameter sensitivities and incorporate cases where there are
integral terms in the objective function, we can extend the original DAE
\eqref{eq:ode_def} by augmenting the state vector with the parameters and the
integrand in the objective function \eqref{eq:int:form} and obtain a larger
system,
\begin{equation}
\label{eqn:ext_sys}
 \ushort{\M} \dot{\ushort{\bu}} = \ushort{F}(t,\ushort{\bu}), ~ t\in [t_0,t_F]\,,
\end{equation}
where
\[
  \setlength{\arraycolsep}{2.5pt}
  \ushort{\M}  =  \left[\begin{array}{ccc} \M & & \\ & \bm{I}_{N_p\times N_p} &  \\ & & 1\end{array}\right],  \ushort{\bu} = \vect{\bu\\\bp\\q},  \ushort{F} = \vect{F\\ \bm{0}_{N_p \times 1}\\r} .
\]
The second equation enforces constant parameters during the time integration,
and the last equation results from a transformation of the integral
\eqref{eq:int:form}.

In this extended framework, the initial condition is $\ushort{\boldeta}_0 = [\boldeta ~ \bp
~ 0]^T$. The extended Jacobian is
\[
  \setlength{\arraycolsep}{2.5pt}
  \ushort{F}_{\ushort{\bu}}  =  \left[\begin{array}{lll} \fu & \fp & \bm{0}_{N_d \times 1} \\ \bm{0}_{N_p \times N_d} & \bm{0}_{N_p \times N_p} & \bm{0}_{N_p \times 1}  \\ r_{\bu} & r_{\bp} & 0 \end{array}\right],
\]
and the extended forward sensitivity matrix is given by
\[
  \setlength{\arraycolsep}{3pt}
  \frac{d \ushort{\bu}}{d \ushort{\boldeta}}  =  
  \left[\begin{array}{lll}
  \frac{d \bu}{d \boldeta} & \frac{d \bu}{d \bp} & \bm{0}_{N_d \times 1} \\
  \bm{0}_{N_p \times N_d} & \bm{I}_{N_p \times N_p} & \bm{0}_{N_p \times 1}  \\
  \frac{d q}{d \boldeta} & \frac{d q}{d \bp} & 0
  \end{array}\right] .
\]
 
The first-order adjoint variable expands to the combination of three variables,
corresponding to the partial derivative of the objective function with respect
to the initial condition of the system state, the parameters, and the initial
value of $q$, respectively. The third variable has a constant value of $1$ because
of the zeros in the last column of $\ushort{F}_{\ushort{\bu}}$ (see the Appendix
in \cite{Zhang2014}).

\subsection{Example: theta methods\label{sec:Theta:method:example}}
As an illustrative example, we describe how the TLM and the first-order and
second-order adjoint models are derived for theta methods, which can be written
as
\begin{equation}
  \M \bu_{n+1} = \M \bu_n + h_n  (1-\theta) \f(\bu_n) + h_n \theta \f(\bu_{n+1})\,,
\label{eqn:theta}
\end{equation}
where $h_n=t_{n+1}-t_{n}$.
%
% A special case Backward Euler reads
% %
% \begin{equation}
%   \M \bu_{n+1} = \M \bu_n + h_n f(\bu_{n+1};\bp).
%   \label{eqn:beuler}
% \end{equation}
% %

\subsubsection{First-order adjoint sensitivity}
In its simplest form, the adjoint theta method for computing solution
sensitivity is
\begin{subequations}
  \begin{align}
    \M^T\blambda_s &= \blambda_{n+1} + h_n\theta\,\fu^T(\bu_{n+1})\blambda_s   \label{eqn:disadj_theta_simple_a}\\
    \blambda_n &= \M^T\blambda_s + h_n (1-\theta)\fu^T(\bu_n)\blambda_s
    \label{eqn:disadj_theta_simple_b}
  \end{align}
  \label{eqn:disadj_theta_simple}
\end{subequations}
with the terminal condition
\begin{equation}
  \blambda_{N} = \left(\frac{\partial \psi}{\partial
    \bu}(\bu_n)\right)^T.
 \end{equation}

By applying this formula to the augmented system \eqref{eqn:ext_sys}, we obtain a method that can compute parameter sensitivities and can incorporate integrals in the objective function:
\begin{equation}
\begin{aligned}
\label{eqn:disadj_theta_full}
  \M^T\blambda_s &= \blambda_{n+1} + h_n\theta\,\fu^T(\bu_{n+1})\blambda_s + h_n \theta\,r_{\bu}^T(t_{n+1},\bu_{n+1}), \\
  \blambda_n  &= \M^T\blambda_s + h_n(1-\theta)\fu^T(\bu_n)\blambda_s + h_n (1 - \theta)r_{\bu}^T(t_n,\bu_n), \\
  \bmu_n &= \bmu_{n+1} + h_n\theta\left(\fp^T(\bu_{n+1})\blambda_s+r_p^T(\bu_{n+1})\right) + h_n(1-\theta)\left(\fp^T(\bu_n)\blambda_s+ r_
  {\bp}^T(\bu_n)\right),
\end{aligned}
\end{equation}
where $\bmu_n = \frac{\partial\psi}{\partial \bp}(\bu_n)$,
$\mathbf{f}_{\{\bu,\bp\}}=\frac{\partial \mathbf{f}}{\partial \{\bu,\bp\}}$, and
$r_{\{\bu,\bp\}}=\frac{\partial r}{\partial \{\bu,\bp\}}$. The corresponding
terminal conditions are
\begin{equation}
 \label{eqn:tcondition_theta}
 \blambda_{N} = \left(\frac{\partial \psi}{\partial
   \bu}(\bu_n)\right)^T, \quad \bmu_{N} = \left( \frac{\partial
   \psi}{\partial \bp}(\bu_n)\right)^T.
\end{equation}

\subsubsection{First-order forward sensitivity}
We take the derivative of the one-step time integration algorithm
\eqref{eqn:theta} with respect to parameters $\bp\in \R^{N_p}$ and obtain the
discrete TLM
\begin{equation}
  \label{eqn:disfwd_theta}
\begin{aligned}
 \M\cS_{n+1} & = \M\cS_{n} + h_n\big( (1-\theta)\left( \fu (\bu_n) \cS_{n} + \fp (\bu_n)\right) \\
 & + \theta \left( \fu (\bu_{n+1}) \cS_{n+1}+\fp(\bu_{n+1}) \right)  \big) \,,
\end{aligned}
\end{equation}
where $\cS_{n} = d \bu_n /d\bp$ denotes the solution sensitivities (a.k.a.
trajectory sensitivities). % It is a matrix and each column is computed
%independently.

With the solution sensitivities, the total derivative of $\psi(\bu_n)$ can be
computed by using
\begin{equation}
\label{eqn:total_der}
\frac{d \psi}{d \bp}(\bu_n) =  \frac{\partial \psi}{\partial \bu}(\bu_n) \cS_{N}  + \frac{\partial \psi}{\partial \bp}(\bu_n)
\end{equation}
or in column-vector form
\begin{equation}
\label{eqn:total_der_cv}
\left(\frac{d \psi}{d \bp}(\bu_n)\right)^T =  \cS_N^T \left(\frac{\partial \psi}{\partial \bu}(\bu_n)\right)^T + \left(\frac{\partial \psi}{\partial \bp}(\bu_n)\right)^T.
\end{equation}

Sensitivity for the integral representation of the objective function is given
by
\begin{equation}
  \label{eqn:total_der_p}
  \frac{d q}{d \bp} = \int_{t_0}^{t_F} \left( \frac{\partial r}{\partial \bu}(\bu) \cS + \frac{\partial r}{\partial \bp} (\bu) \right) \, dt.
\end{equation}
%

% \subsection{Second-order FOF}
% For simplicity, first we consider the sensitivities to initial conditions. 
% \begin{multline}
%  \label{eqn:disfwd_theta_fof}
%  \M \cS_{\ell, n+1}^{(2)} = \M \cS_{\ell, n}^{(2)} + h_n \big( (1-\theta)\left( \fuu (\bu_n) \cS_{\ell, n}^{(1)} \cS_{\ell, n}^{(1)} + \fu (\bu_n) \cS_{\ell, n}^{(2)} \right) \\
%  + \theta \left(\fuu (\bu_{n+1}) \cS_{\ell, n+1}^{(1)} \cS_{\ell, n+1}^{(1)} + \fu (\bu_{n+1}) \cS_{\ell, n+1}^{(2)} \right)  \big) .
% \end{multline}

\subsubsection{Second-order adjoint: sensitivities to initial condition}
Differentiating the first-order adjoint \eqref{eqn:disadj_theta_simple} with
respect to the initial condition leads to
\begin{subequations}
\label{eqn:2nddisadj_theta}
  \begin{align}
  %\nonumber
  \M^T \, \frac{d \blambda_s}{d \boldeta} &= \frac{d \blambda_{n+1}}{d \boldeta} + h_n \theta  \blambda_{s}^T \fuu(\bu_{n+1}) \, \frac{d \bu_{n+1}}{d \boldeta}  + h_n \theta \fu^T(\bu_{n+1}) \, \frac{d \blambda_s}{d \boldeta} \\
  %\nonumber
  \frac{d \blambda_n}{d \boldeta}  &= \M ^T \, \frac{d \blambda_s}{d \boldeta} + h_n (1-\theta) \blambda_s^T \fuu(\bu_n) \, \frac{d \bu_n}{d \boldeta} + h_n (1-\theta) \fu^T(\bu_n) \, \frac{d \blambda_s}{d \boldeta},
%  \nonumber
%  \bmu_{\ell,n} &= \bmu_{\ell,n+1} + h_n \left(\fup^T(\bu_{n+1}) \cS_{\ell,n} + \fppl^T(\bu_{n+1})\right)\blambda_s \\
%            &+ h_n \left( r_{\bu\bp}^T(t_{n+1},u_{n+1}) \cS_{\ell,n} + r_{pp_\ell}^T(t_{n+1},u_{n+1}) \right)
\end{align}
\end{subequations}
with the terminal condition
\begin{equation} 
\label{eqn:tcondition2}
  \frac{d \blambda_{N}}{d \boldeta} = \frac{d}{d \bu}\left(\frac{d \psi}{d \bu}(\bu_n)\right)^T \frac{\partial \bu_n}{\partial \boldeta}.
\end{equation}
Post-multiplying both sides of \eqref{eqn:2nddisadj_theta} by a direction vector
$\bv \in \R^{N_d}$ and defining $\bLambda = (d \blambda / d \boldeta) \bv$ to
shorten the expression, we obtain
\begin{equation}
\label{eqn:2nddisadj_theta_mf}
  \begin{aligned}
    \M^T\bLambda_s &= \bLambda_{n+1} + h_n\theta\blambda_s^T\,\fuu(\bu_{n+1}) \boxed{\frac{d \bu_{n+1}}{d \boldeta}\bv} + h_n\theta\fu^T(\bu_{n+1}) \bLambda_s \\
    \bLambda_n &= \M^T\bLambda_s + h_n(1-\theta)\blambda_s^T \fuu(\bu_n)\boxed{\frac{d \bu_n}{d \boldeta} \bv} + h_n(1-\theta)\fu^T(\bu_n) \bLambda_s
  \end{aligned}
\end{equation}
with the terminal condition
\begin{equation}
  \label{eqn:tcondition}
  \bLambda_N = \frac{d}{d \bu}\left(\frac{d \psi}{d \bu}(\bu_n)\right)^T \frac{\partial \bu_n}{\partial \boldeta} \bv.
 \end{equation}

Comparing the second-order adjoint \eqref{eqn:2nddisadj_theta_mf} with the
first-order adjoint \eqref{eqn:disadj_theta_simple}, one can see that they are
similar; the only difference is the additional term containing the
Hessian-vector product of the DAE right-hand side. They result in linear systems
with the same shifted Jacobian matrix $\M^T- h_n \theta \fu^T(\bu_{n+1})$ but
different right-hand sides. Therefore, they can be solved together with those in
the first-order adjoint, using the same preconditioners.

For large-scale simulations, computing the full forward sensitivity matrix
$\frac{d \bu_n}{d \boldeta}$ quickly becomes impractical because it requires a
computational cost that is linear with the number of states. However,
calculating the directional derivatives for the forward sensitivities (boxed
terms in \eqref{eqn:2nddisadj_theta_mf}) makes the cost constant; as a result,
the computational cost of the second-order adjoint is independent of the number of
inputs (states and parameters), like the cost of the first-order adjoint.

% The Hessian-vector product of the DAE right-hand side can be generated
% efficiently by algorithmic differentiation tools or approximated by Jacobian
% evaluations. 

\subsubsection{Second-order adjoint: sensitivities to parameters}
We can apply techniques similar to those described in Section
\ref{sec:augmented_sys} to extend the method for computing solution
sensitivities to cases where parameter sensitivities are desired and integrals
are included in the objective function.

The extended Hessian of the DAE right-hand side contains $3 \times 3 \times 3$
tensor blocks, including $\fuu$, $F_{\bu\bp}$, $F_{\bp\bu}$, $F_{\bp\bp}$,
$r_{\bu\bu}$, $r_{\bu\bp}$, $r_{\bp\bu}$,  and $r_{\bp\bp}$, and 19 zero blocks. The
vector-Hessian product term in \eqref{eqn:2nddisadj_theta}, $\blambda_s^T
\fuu(\bu_n)$, is
\[
  \setlength{\arraycolsep}{3pt}
  \left[\begin{array}{lll}
  \blambda^T \fuu + r_{\bu\bu} & \blambda^T \fup + r_{\bu\bp} & \bm{0} \\
  \blambda^T \fpu + r_{\bp\bu} & \blambda^T \fpp + r_{\bp\bp} & \bm{0} \\
  0 & 0 & 0
  \end{array}\right] .
\]

We also need to extend the second-order adjoint variable multiplied with a
directional vector to three variables denoted by $\bLambda$, $\bGamma$, and
$\Theta$. The corresponding directional vector should be split into three
components $\bv_1 \in \R^{N_d}$, $\bv_2 \in \R^{N_p}$, and $\bv_3 \in \R^{1}$. We define the new directional forward
sensitivity to be $\bw_1 \in \R^{N_d}, \bw_2 \in \R^{N_p}$, and $\bw_3 \in \R^{1}$  for the boxed term in
\eqref{eqn:2nddisadj_theta_mf}, where
\[
  \vect{\bw_1(\bu_n)\\\bw_2(\bu_n)\\\bw_3(\bu_n)} = \frac{d \ushort{\bu_n}}{d \ushort{\boldeta}} \vect{\bv_1\\\bv_2\\\bv_3}.
\]

Multiplying the vector-Hessian product term with the directional forward
sensitivities eliminates $\bw_3$ because of the zeros in the last row and leads
to $\bw_2=\bv_2$ because of the identity in the center. Thus, only $\bw_1$ needs
to be obtained by solving the TLM equation
\begin{equation}
\begin{aligned}
 \label{eqn:fwdsen_w}
 \M \bw_{n+1} & = \M \bw_n + h_n\big( (1-\theta)\left( \fu (\bu_n) \bw_n + \fp(\bu_n) \bv_2 \right) \\
 & + \theta \left( \fu (\bu_{n+1}) \bw_{n+1}+\fp(\bu_{n+1}) \bv_2 \right)  \big).
\end{aligned}
\end{equation}
See the supplementary material for details.

Expanding the augmented system leads to
\begin{equation}
  \begin{aligned}
  \M^T \, \bLambda_s &= \bLambda_{n+1} + h_n\theta\,\fu^T(\bu_{n+1})\,\bLambda_s \\
  & + h_n\theta\left(\blambda_s^T\,\fuu(\bu_{n+1})\bw_1(\bu_{n+1})+r_{\bu\bu}(\bu_{n+1})\bw_1(\bu_{n+1})\right)\\
  & +h_n\theta\left(\blambda_s^T\,\fup(\bu_{n+1})\bw_2(\bu_{n+1}) + r_{\bu\bp}(\bu_{n+1})\bw_2(\bu_{n+1})\right) \\
  \bLambda_n &= \M^T\,\bLambda_s +  h_n(1-\theta)\fu^T(\bu_n)\bLambda_s \\
  & + h_n(1-\theta)\left(\blambda_s^T\,\fuu(\bu_n)\bw_1(\bu_n) +r_{\bu\bu}(\bu_n)\bw_1(\bu_n)\right)\\
  & + h_n(1-\theta)\left(\blambda_s^T\,\fup(\bu_n)\bw_2(\bu_n) +r_{\bu\bp}(\bu_n)\bw_2(\bu_n)\right) \\
  \bGamma_n &= \bGamma_{n+1} + h_n\theta \fp^T(\bu_{n+1})\,\bLambda_s  \\
  & + h_n\theta\left(\blambda_s^T\,\fpu(\bu_{n+1})\bw_1(\bu_{n+1})+r_{\bp\bu}(\bu_{n+1})\bw_1(\bu_{n+1})\right) \\ 
  & + h_n\theta\left(\blambda_s^T\,\fpp(\bu_{n+1})\bw_2(\bu_{n+1}) + r_{\bp\bp}(\bu_{n+1})\bw_2(\bu_{n+1})\right) \\
  & + h_n(1-\theta)\fp^T(\bu_n)\,\bLambda_s \\
  & + h_n(1-\theta)\left(\blambda_s^T\,\fpu(\bu_n)\,\bw_1(\bu_n)+r_{\bp\bu}(\bu_n)\bw_1(\bu_n)\right)\\
  & + h_n(1-\theta)\left(\blambda_s^T\,\fpp(\bu_n)\,\bw_2(\bu_n) + r_{\bp\bp}(\bu_n)\bw_2(\bu_n)\right)\\
  \Theta_n &= \Theta_{n+1}
  \end{aligned}
\end{equation}
with terminal conditions 
\begin{equation}
  \begin{aligned}
  \bLambda_N &= \DDPSI{\bu}{\bu}{(\bu_n)}\frac{\partial \bu_n}{\partial \boldeta} \bv_1 + \left(\DDPSI{\bu}{\bu}{(\bu_n)}\frac{\partial \bu_n}{\partial \bp}+\DDPSI{p}{\bu}{(\bu_n)}\right) \bv_2 \\
  \bGamma_N &= \DDPSI{\bu}{p}{(\bu_n)}\frac{\partial \bu_n}{\partial \boldeta}\bv_1 + \left(\DDPSI{\bu}{p}{(\bu_n)}\frac{\partial \bu_n}{\partial \bp}+\DDPSI{p}{p}{(\bu_n)}\right)\bv_2.
\end{aligned}
\end{equation}
The final solution is given by
\begin{equation}
  \begin{aligned}
  \bLambda_0 =  \DDPSI{\boldeta}{\boldeta}{}\bv_1 + \DDPSI{\bp}{\boldeta}{}\bv_2 \\
  \bGamma_0 =  \DDPSI{\boldeta}{\bp}{}\bv_1 + \DDPSI{\bp}{\bp}{}\bv_2 .
\end{aligned}
\end{equation}

To compute the total derivatives for $\psi$, we can apply the chain rule with
the adjoint solution
\begin{equation}
  \label{eq:tot:phi:optimization}
  \nabla_{\bp}\psi = \left(\frac{d \psi}{d \bp}\right)^T =  \left(\frac{d \boldeta}{d \bp}\right)^T\left(\frac{\partial \psi}{\partial \boldeta}\right)^T+ \left(\frac{\partial \psi}{\partial \bp}\right)^T = \boldeta_p^T\blambda_0 + \bmu_0.
\end{equation}
Similarly, the second-order directional derivative with respect to the
parameters can be computed as
\begin{equation}
  \begin{aligned}
  \nabla_{\bp}^2\psi\,\bsigma &= \frac{d }{d \bp}\left(\frac{d \psi}{d \bp}\right)^T\,\bsigma \\
  &= \frac{\partial \psi}{\partial \boldeta}\boldeta_{\bp\bp}\,\bsigma+ \boldeta_{\bp}^T\left(\DDPSI{\boldeta}{\boldeta}{}\boldeta_{\bp}+\DDPSI{\bp}{\boldeta}{}\right)\bsigma+\DDPSI{\boldeta}{\bp}{}\boldeta_{\bp}\,\bsigma\\
  &+\DDPSI{\bp}{\bp}{}\bsigma\\
  &= \blambda_0^T\,\boldeta_{\bp\bp}\,\bsigma + \boldeta_{\bp}^T\,\bLambda_0+\bGamma_0
  \end{aligned}
\end{equation}
with $\bv_1=\boldeta_{\bp}\bsigma$ and $\bv_2=\bsigma$. At this point, the second-order
derivative with respect to the initial conditions is simply 
\begin{equation}
  \begin{aligned}
  \nabla_{\boldeta}^2\psi\,\bsigma &= \frac{d }{d \boldeta}\left(\frac{d \psi}{d \boldeta}\right)^T\,\bsigma = \bLambda_0 \, ,
  \end{aligned}
\end{equation}
with $\bv_1=\bsigma$.

\section{PETSc TSAdjoint}
We begin this section with an overview of the  \texttt{PETSc}
\texttt{TSAdjoint} software and then discuss the design and user interface as well as
some implementation issues.

\subsection{Overview of the software}
\texttt{PETSc} is a scalable MPI- and GPU-based object-oriented numerical
software library written in C and fully usable from C, C++, Fortran, and Python.
It is publicly available at \url{https://www.mcs.anl.gov/petsc/}. \texttt{PETSc}
has several fundamental classes from which applications are composed, including
data structures for vectors and matrices, abstractions for working with
subspaces of vectors, linear and nonlinear solvers, ODE/DAE solvers, and
optimization solvers (within the Toolkit for Advanced Optimization
(\texttt{TAO}) component of \texttt{PETSc}). In addition, \texttt{PETSc} has an
abstract class DM that serves as an adapter between meshes, discretizations, and
other problem descriptors and the algebraic and timestepping objects that are
used to solve the discrete problem.
% In addition
% \texttt{PETSc} provides both a programmatic interface to almost all functionalities
% within \texttt{PETSc} and a simple string-based system, called the options database, that
% allows runtime control of almost all functionality in PETSc.

\texttt{PETSc} \texttt{TSAdjoint} provides a number of advantages. It avoids the
full differentiation of a simulation code that classic AD requires, while
maintaining the accuracy and speed of using AD tools. \texttt{PETSc} also offers
finite-difference approximations for validating the user-supplied Jacobian (or
Jacobian-vector products in a matrix-free context) and even the adjoint
sensitivities.
%\texttt{PETSc} also offers
%finite-difference approximations for gradient computations, which can be used to
%generate Jacobian matrices (or Jacobian-vector products in a matrix-free
%context), as well as to validate the user-supplied Jacobian and even the adjoint
%sensitivities.
Users can easily enable these functionalities via command-line options at
runtime. Compared with the continuous adjoint approach \cite{Jameson1988} that
has been popular in control theory for a long time, the discrete adjoint
approach adopted in \texttt{PETSc} does not require users to derive a new set of
PDEs and determine boundary conditions to ensure the existence of the solution
of the adjoint equations. One may argue that the continuous adjoint approach
allows different discretization schemes and adaptive techniques to be applied to
the adjoint equation, giving opportunities for efficiency improvement. While
this is reasonable in theory,  implementation and accuracy concerns
 may arise in applications. First, adapting spatial discretization is not
trivial, since it may involve changes to the mesh and need extra code
development to implement the new schemes. Second, interpolation in the temporal
domain becomes necessary when the checkpointed data from the original forward
model cannot be used directly in solving the continuous adjoint equation (e.g.,
when adaptive timestepping is enabled). The interpolation will also induce
additional numerical errors. Thus, exploiting the flexibility in choosing
discretization schemes for continuous adjoint approaches can be a tremendous
burden for application developers. Third, it has been shown that the discrete
adjoint approach can deliver better accuracy than can the continuous adjoint
approach in machine learning \cite{Onken2020,Gholami2019}. The abstraction level
at which the discrete adjoint model in \texttt{PETSc} is derived provides a
balance between flexibility and usability---it does not raise concerns about
discretization, and it still offers flexibility in the selection of algebraic
solvers.

Various checkpointing schemes have been implemented in a new class
called \texttt{TSTrajectory}, which generates an optimal checkpointing schedule
used internally by \texttt{TSAdjoint}, thus being completely transparent to
users. Using an optimal checkpointing schedule is critical for achieving good
performance in adjoint calculations. It is a difficult combinatorial problem and
orthogonal to the focus of application developers. Therefore, the implementation
of automatic checkpointing is a significant advantage to application developers.

\subsection{Design and user interface}
Rooted in the \texttt{PETSc} timestepping library \cite{Abhyankar2018},
\texttt{TSAdjoint} is designed
for the scalable computation of sensitivities of systems of
time-dependent PDEs, DAEs, and ODEs. For each class of time integration methods
in PETSc, a corresponding adjoint version of the algorithm is implemented with
the context (e.g., method coefficients, working vectors) shared with the forward
timestepping solver. The adjoint solvers are provided with event detection and
handling (\texttt{TSEvents}), solution monitoring (\texttt{TSMonitor}), and
performance profiling and thus are feature-complete compared with their
counterparts. The event feature is particularly crucial for handling hybrid
dynamical systems with discontinuities (or jumps) in time. These problems are
known to be challenging for sensitivity analysis because complicated jump
conditions at the switching surface need to be derived and implemented.
Interested readers can refer to \cite{Zhang2017} for details on how this
capability is achieved with \texttt{PETSc} \texttt{TSAdjoint}.

In \texttt{PETSc}, DAEs and ODEs are formulated as $\F(t,\bu,\dot{\bu})=
\G(t,\bu)$. For clarity of presentation, the form considered in this paper,
\eqref{eq:ode_def}, is a common case where $\F = \M \dot{\bu} - \f$ and
$\G=\bm{0}$; but \texttt{TSAdjoint} is extensible to fully support the more
general case. To utilize the \texttt{PETSc} integrators, users supply callback
routines for the residual function ($\F$ and $\G$) evaluations and optional
routines for Jacobian evaluation when implicit methods are chosen. For example,
the Jacobian with respect to the state for \eqref{eq:ode_def}, by the chain
rule, is $a\F_{\dot{\bu}} + \Fu$, where the shift parameter $a$ depends on the
time integration method and is passed to the user's callback routine. For
sensitivity analysis, these same callbacks are reused, but a few additional
callbacks may be required to provide derivatives (Jacobian and Hessian) of the
ODE/DAE operator with respect to system state or parameters depending on the
application needs. The Jacobian can be given either directly or in a matrix-free
form. The matrix-free form (vector-Hessian-vector product) is preferred for the
Hessian because the sensitivity analysis techniques do not need to use the
matrix or tensor directly, the memory footprint can be dramatically reduced, and
the vector-Hessian-vector product can be generated much more efficiently by AD
tools than can the Hessian itself. The vectors to be multiplied with the Hessian
are also prepared by \texttt{PETSc} and accessed by users through the API. Table
\ref{tab:callbacks} summarizes the callback routines for several typical use
cases.
\def\arrvline{\hfil\kern\arraycolsep\vline\kern-\arraycolsep\hfilneg}
\begin{table}
  \centering
  \caption{\label{tab:callbacks}User-supplied callbacks for an implicit timestepping solver and its adjoint calculations. Reusable callbacks across use cases are marked in gray.}
  \renewcommand{\arraystretch}{1.1} 
    \begin{tabular}{l  c  c  c  c c c}
      \hline
      Use case & \multicolumn{3}{c}{Without integral} & \multicolumn{3}{c}{With integral } \\
      \hline
      \multirowcell{2}{forward integration} & \multirowcell{2}{$\M \dot{\bu} - \f$\\ $ a\M - \fu$} & & & \multirowcell{2}{$r$} & & \\ \\
      \hline
      \multirowcell{2}{1st-order adjoint \\ or TLM} & \textcolor{gray}{  \multirowcell{2}{$\M \dot{\bu} - \f$\\ $ a\M - \fu$} } &  \multirowcell{2}{$-\fp$} &  & \textcolor{gray}{ \multirowcell{2}{$r$} }& \multirowcell{2}{$r_{\bu}$ \\ $r_{\bp}$} & \\ \\
      \hline
      \multirowcell{4}{2nd-order adjoint} & \textcolor{gray}{  \multirowcell{4}{$\M \dot{\bu} - \f$\\ $ a\M - \fu$} }& \textcolor{gray}{ \multirowcell{4}{$-\fp$} }&  \multirowcell{4}{$-\bv_1^T \fuu \bv_2 $ \\ $-\bv_1^T \fup \bv_2 $\\ $-\bv_1^T \fpu \bv_2 $\\ $-\bv_1^T \fpp \bv_2 $ } &  \textcolor{gray}{ \multirowcell{4}{$r$} }& \textcolor{gray}{ \multirowcell{4}{$r_{\bu}$ \\ $r_{\bp}$} }& \multirowcell{4}{$r_{\bu\bu} \bv_3$\\ $r_{\bu\bp} \bv_3 $ \\ $r_{\bp\bu} \bv_3$\\ $r_{\bp\bp} \bv_3 $} \\ \\ \\ \\
      \hline
    \end{tabular}
\end{table}

The user interface to the adjoint solver is consistent with that of the
timestepping solver. In particular, users need to create the appropriate
\texttt{PETSc} vectors for storing the adjoint variables, provide the
problem-specific context using \texttt{TSSetCostGradients()} for first-order
adjoints, and initialize the adjoint variables according to the proper terminal
conditions between the end of the forward solve and the start of the adjoint
solve. For the second-order adjoint, additional adjoint variables need to be
provided using \texttt{TSSetCostHessianProducts()}, and tangent linear variables
need to be set with \texttt{TSAdjointSetForward()}.

% timestep adaptivity
Adaptive timestepping is naturally supported. Both the tangent linear and
adjoint solvers follow the same trajectory that the
timestepping solver determines via a timestep controller. The \texttt{PETSc} timestepping
solver provides a variety of options for automatic timestep control to
attain a user-specified goal. The adaptivity logic can be based on embedded
error estimates \cite{Dormand1980}, linear digital control theory
\cite{Soderlind2003}, the Courant--Friedrichs--Lewy condition, and global
error estimates \cite{Constantinescu2018}. When adaptive timestepping is used, an
online checkpointing scheme must be employed because the total number of steps
is not known a priori.

\subsection{Jacobian/Hessian computation}
\texttt{PETSc} provides several choices for the Jacobian/Hessian operators or
their application needed by the forward and adjoint solvers. First,
\texttt{PETSc} offers efficient and automatic Jacobian approximation with finite
differences and coloring \cite{Gebremedhin2005} if the Jacobian is not supplied
by users and the sparsity pattern of the Jacobian is available (e.g., when the
\texttt{PETSc} data management object \texttt{DM} is used for the implementation
of discretization schemes). Second, \texttt{PETSc} allows low-level AD tools to
differentiate local routines so that MPI routines need not be differentiated
through, and it provides utilities to facilitate fast Jacobian recovery from
AD-generated matrices (see \cite{wallwork2019} for details). Third, one can use
libraries such as Firedrake and FEniCS that have excellent high-level AD
capabilities; this use is demonstrated with examples in Section \ref{sec:examples}.

\section{Checkpointing}

\begin{figure}
  \begin{subfigure}{1.0\linewidth}
    \centering
    \resizebox{0.9\linewidth}{!}{
    \includegraphics{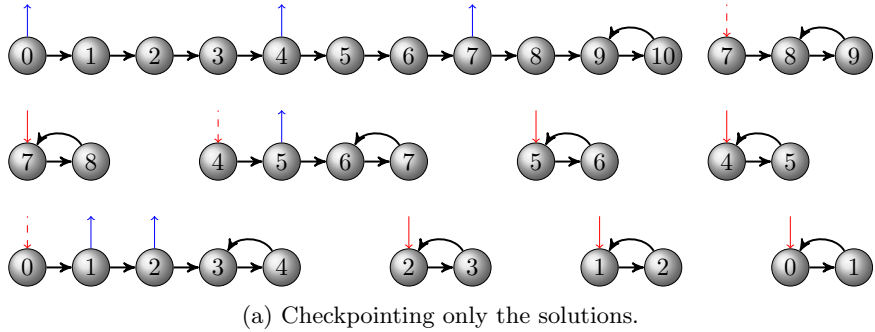}
    }
  \caption{Checkpointing only the solutions.}
  \label{fig:process_a}
  \end{subfigure}
  \begin{subfigure}{1.0\linewidth}
    \centering
    \resizebox{0.9\linewidth}{!}{
    \includegraphics{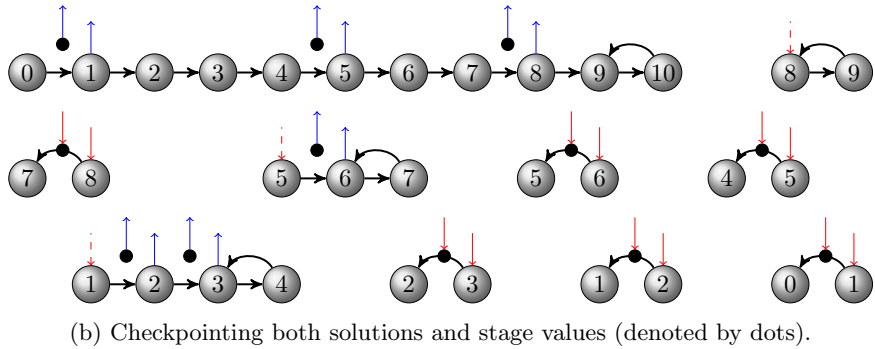}
    }
  \caption{Checkpointing both solutions and stage values (denoted by dots).}
  \label{fig:process_b}
  \end{subfigure}
  \caption{From left to right, top to bottom: the processes controlled by (a) \texttt{revolve} and (b) modified \texttt{revolve}. Numbered nodes stand for solutions at each time step.} The up arrow and down arrow stand for ``store'' operation and ``restore'' operation, respectively. When a stack is used for holding the checkpoints, the arrows with solid lines correspond to push and pop operations. The down arrow with dashed line indicates  reading the top element on the stack without removing it.
  \label{fig:process}
\end{figure}

In order to calculate the discrete adjoint state, Jacobians and Hessians or
matrix-free operations for them must be evaluated by using the system states
that are computed in the forward run. The storage space needed to retain all
these states is proportional to the number of time steps performed. To overcome
this drastic storage requirement, one can checkpoint selective states along the
trajectory while recomputing the missing ones. This technique has been well
studied in the literature. A notable offline algorithm, \texttt{revolve},
developed by Griewank and Walther \cite{Griewank2000}, generates a checkpointing
schedule that minimizes the number of recomputation time steps, given the total
number of time steps and the number of allowed checkpoints in memory. A C++ tool
was developed to implement the \texttt{revolve} algorithm; a few online
algorithms \cite{Heuveline2006,Stumm2010,Wang2009} were also implemented for
cases when the number of time steps is not known a priori; and a multistage
algorithm was included to consider both disk and memory for storage
\cite{Stumm2009}. Figure \ref{fig:process_a} depicts an optimal schedule for
adjoining $10$ time steps given three checkpoints.

However, using these algorithms and the tool can cause difficulties. First, they
provide only the schedule that guides the checkpoint manipulation for adjoint
computation. Significant effort is still needed to implement the required
operations that are dependent on the application codes and hardware platforms.
% For example, relevant questions include how to move the data to the designated
% storage media and in which format and how to change the workflow so that time
% steps could be recomputed between checkpoint access and adjoint state
% calculation.
Second, the tool was designed to be an explicit controller for conducting
forward integration and adjoint integration in time-dependent applications. Most
ODE solvers, however, have their own framework for controlling the timestepping
process. Incorporating \texttt{revolve} in these software systems can be intrusive or
even infeasible. For example, the adjoint solve involves a workflow that mixes
forward and reverse integration, which are not commonly supported
in existing ODE solvers. Third, \texttt{revolve} was designed under the
assumption that only solution states are checkpointed at distinct time steps; it
requires at least one recomputation before each adjoint step can be performed.
This strategy is not necessarily ideal for the discrete adjoint of multistage
time integration methods because checkpointing the intermediate stage values
together with the solution states would remove the need to recompute the
corresponding time steps.

To address these challenges, we have implemented the \texttt{TSTrajectory}
component in \texttt{PETSc} to serve as the intermediary between
\texttt{revolve} and the timestepping solver. It is responsible for implementing
the operations required by \texttt{revolve} and handling the adjoint workflow.
The main features are summarized below.
\begin{itemize}
  \item Storing and restoring a checkpoint are implemented for different storage
  media. In memory, these operations are straightforward; on disk or other
  devices, data format and parallel I/O must be considered. For example, we
  currently support binary file formats and MPI I/O, but this support can easily
  be extended to other possibilities.
  \item Needed data points can be requested from \texttt{TSTrajectory} by
  specifying either the timestep number (a unique index for labeling each time
  step) or the time. In the forward run, selected checkpoints will be stored. In
  the reverse run, the data point needed to complete an adjoint step is
  restored directly if it has already been checkpointed; and then the checkpoint
  can be discarded to leave the storage space for a new checkpoint. If not
  available immediately, the data point will be recomputed from the nearest
  checkpoint. During the recomputation, a new checkpoint may be stored if
  storage space permits. This reinterpretation of the checkpointing schedule
  allows us to encapsulate the process of obtaining a data point into
  \texttt{TSTrajectory} and hide it from the requesting code (i.e., the adjoint
  solver).
  \item For multistage time integration methods \texttt{TSTrajectory} allows
  users to checkpoint only the solution or the solution plus the stage values.
  The latter choice may result in further savings in recomputations for some
  cases.
\end{itemize}

Figure \ref{fig:process_b} illustrates (a) an optimal checkpointing schedule
given a storage capacity for $3$ solutions and (b) an optimal checkpointing
schedule, modified from (a), given a storage capacity for $3$ solutions and all
the stage values associated with these solutions. One can see that with a
similar schedule, the first case requires $15$ extra recomputations in the
reverse run while the second case involves $6$ extra recomputations. The
reduction in recomputations results from the fact that a time step can be
directly adjoined for multistage time integration methods if the stage values
are available in memory. 
%More discussions about the modification for \texttt{revolve} and its optimality can be found in the supplementary material. 

%

%

\begin{figure}
  \centering
  \includegraphics[width=0.95\textwidth]{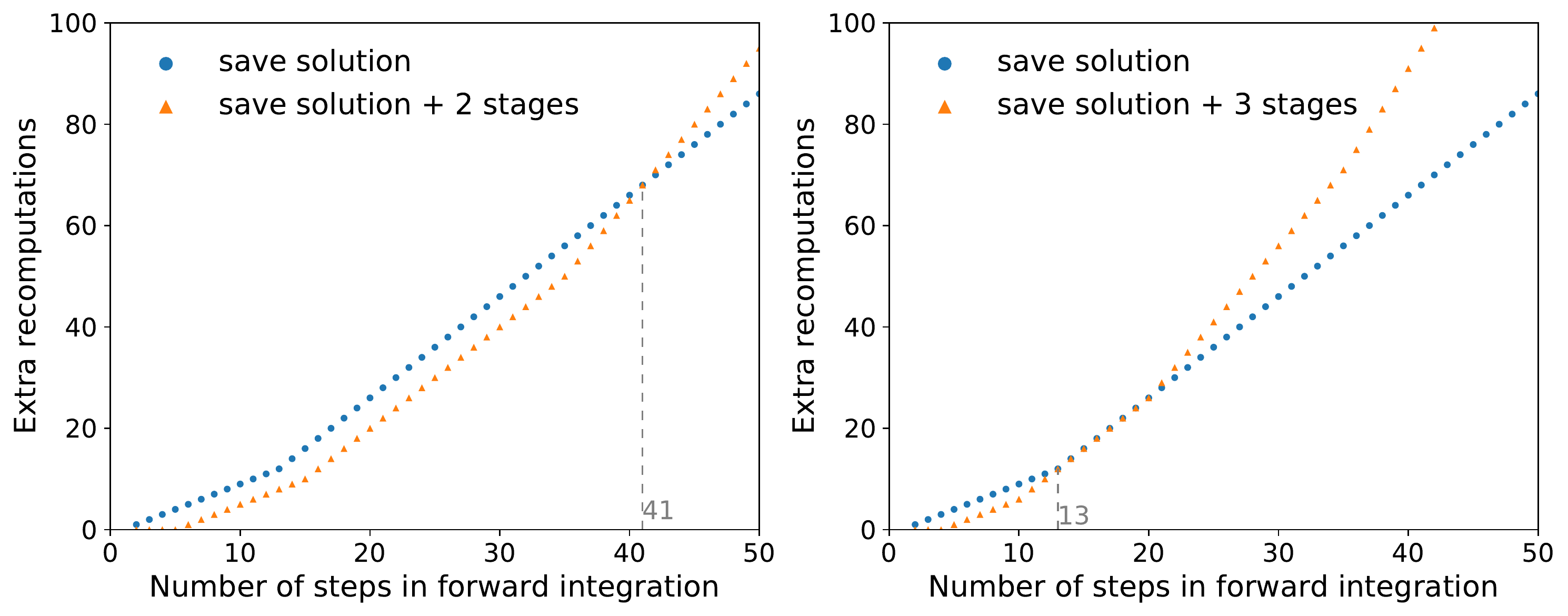}
  \caption{Comparison in terms of recomputations between checkpointing only
 solutions and checkpointing solutions and stage values. In this example we
 assume that the memory available at runtime can hold up to $12$ units (one
 solution or one stage corresponds to one unit). If
 we save only the solution at a time step, $12$ checkpoints can be used. Saving one solution and two stages
 results in $4$ checkpoints available, and saving one solution and three stages
 leads to $3$ checkpoints.}
  \label{fig:mrerolve}
\end{figure}

As a simplistic example to show the potential benefit of the modified
\texttt{revolve} algorithm, we consider the adjoint checkpointing schedule given
a limited amount of memory that can be used to hold $12$ units (one unit
corresponds to one solution or one stage). For a two-stage time integration
method, one can also use the memory to store the data for $4$ time steps ($1$
solution and $2$ stages at each step). Similarly, for a three-stage method, the
data for at most $3$ time steps can be stored in memory. Figure \ref{fig:mrerolve}
illustrates the performance of these options. We observe that saving
only the solutions is not always optimal in terms of extra recomputations and that
saving the stage values along with the solutions, although leading to fewer
``checkpoints'' available, can require fewer recomputations when the total number of
time steps to be adjoined is below certain thresholds.

% Although the workflow is not controlled directly by
% \texttt{revolve}, it is equivalent to the one scheduled by \texttt{revolve}, and
% it allows the timestepping loops in \texttt{TS} and \texttt{TSAdjoint} to be
% unchanged.
%not considering the First Same As Last (FSAL) property 
%

\section{Examples}\label{sec:examples}

This section presents three representative examples from a diverse set of
problems. The goals are to (1) illustrate the use of the \texttt{PETSc}
\texttt{TSAdjoint} in outer-loop applications such as optimal control and
inverse problems, (2) demonstrate the efficiency and scalability of the
implementation, and (3) show the usability of \texttt{PETSc} \texttt{TSAdjoint}
in other scientific computing libraries. To date, \texttt{TSAdjoint} has been
applied in domains including power systems \cite{Zhang2017}, data assimilation
\cite{Carracciuolo2018}, and computational fluid dynamics \cite{Marin2018}.
These applications are not covered in this paper. We refer readers to these
references for more information.

\subsection{An optimal control problem} \label{sec:optimal_control_ex} %Trajectory planning

\begin{figure}
\begin{subfigure}{0.35\linewidth}
\resizebox{\textwidth}{!}{%
\begin{tikzpicture}
  \begin{axis}[
    xmin=0,   xmax=3,
    ymin=0,   ymax=3,
    xticklabels={,,},
    yticklabels={,,}
]
    \addplot[mark=none,red,ultra thick]{x};
	\addplot[color=blue,ultra thick, smooth,dashed
] coordinates {
		(1.5,0)
        (1.65,0.2)
        (1.7,0.45)
        (1.6,0.8)
        (1.63,1.3)
        (1.8,1.7)
        (1.9,1.9)
	};
  \addplot[color=blue,ultra thick, smooth,dashed
] coordinates {
    (1.5,0)
        (1.65,0.2)
        (1.8,0.8)
        (1.9,1.3)
        (2.2,2.1)
  };
	\end{axis}
    \node[color=red,scale=2] at (0,0) {\faPlane};
    \node[color=red,scale=2] at (6.6,5.5) {\faPlane};
    \node[color=blue,scale=2] at (3.44,0) {\faPlane};
    \node[color=blue,scale=2] at (4.3,3.5) {\faPlane};
    \node[color=blue,scale=2,rotate=15] at (5,3.9) {\faPlane};
\end{tikzpicture}
}
\caption{Trajectory of the leader (red) and candidate trajectories of the pursuer (blue).}
\label{fig:aircraft_traj}
\end{subfigure}
~
\begin{subfigure}{0.6\linewidth}
  \centering
  \includegraphics[width=\linewidth]{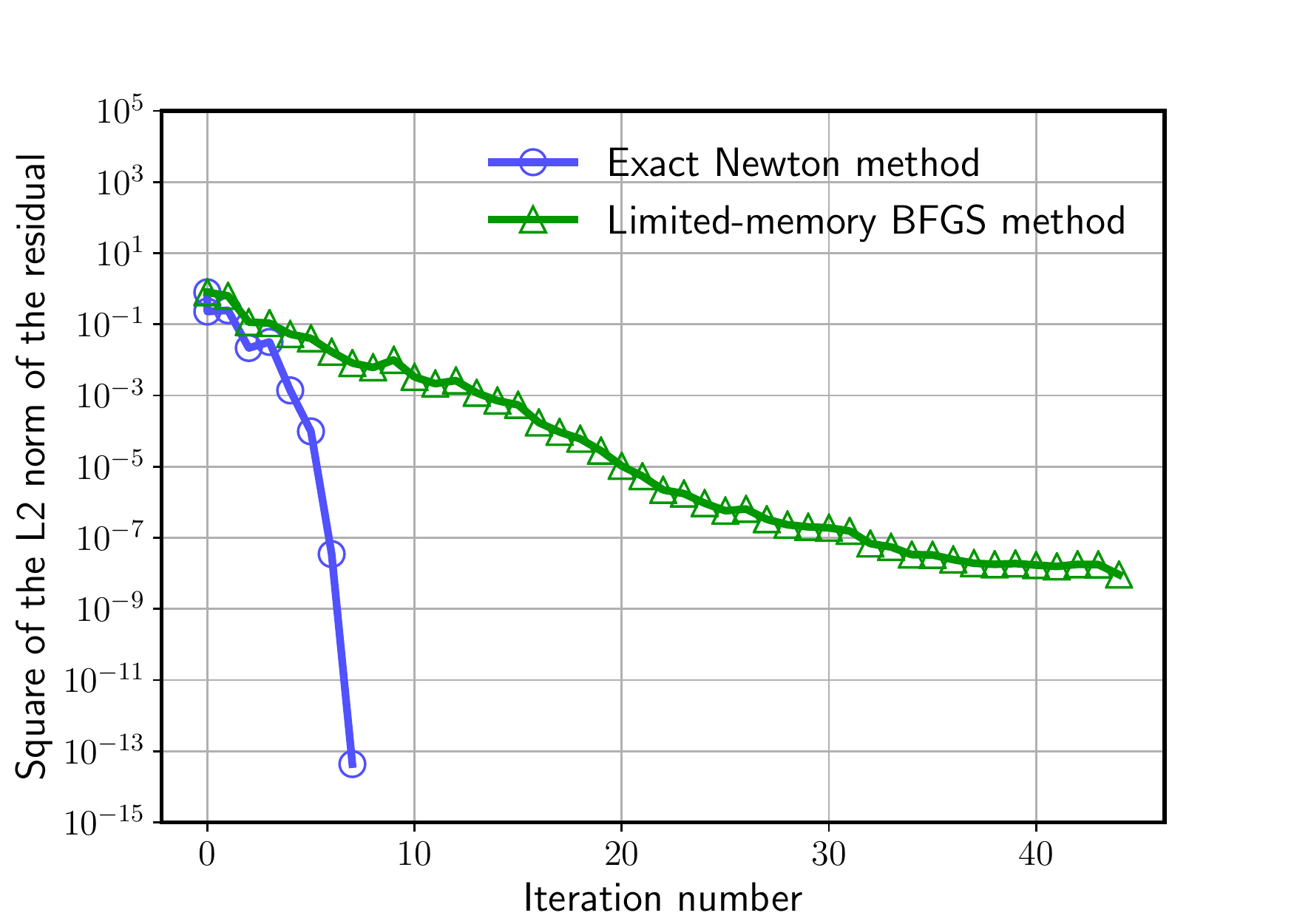}
  \caption{Convergence of the optimization methods.}
  \label{fig:ex1_conv}
\end{subfigure}
\caption{Aircraft trajectory planning: (a) schematic of the problem and (b) comparison in convergence between the limited-memory BFGS method and the Newton method with the exact Hessian in matrix-free form.}
\end{figure}
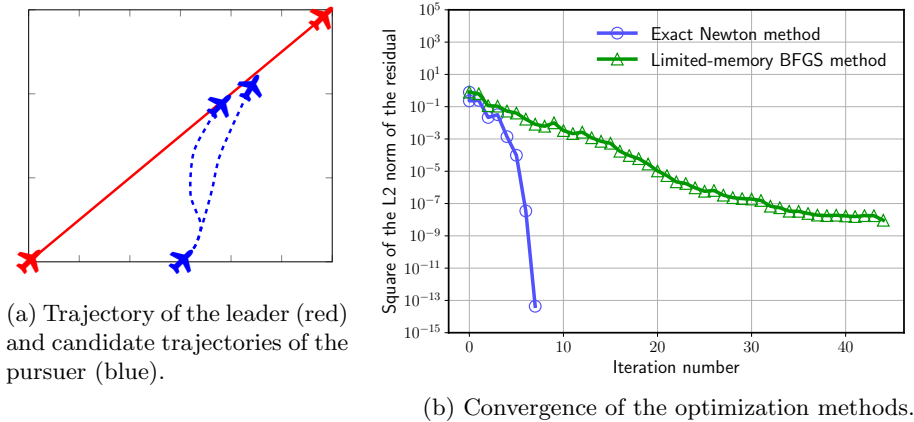

The goal of aircraft trajectory planning is to find a control sequence that can
control the pursuer to the targeting leader by minimizing a given cost function,
as illustrated in Figure \ref{fig:aircraft_traj}. The sequence is divided into
finite time intervals $T_k = [t_k, t_{k+1}]$ for $k =0,1,\dots,N-1$. In each
interval, control inputs are provided in response to the changes in the leader's
position. The dynamics of the aircraft is governed by a nonlinear kinematic
model
\begin{equation}
  \begin{aligned}
    & \dot{x}_k(t) = v_k (t) \cos (\omega_k(t)) \\
    & \dot{y}_k(t) = v_k (t) \sin (\omega_k(t))
  \end{aligned}
  \label{eq:aircraft_dyn}
\end{equation}
defined on each time interval $T_k$.

The problem can be transformed into the minimization of the cost function
\begin{equation}
  \psi (\bu,\bp) = \int_0^{t_F} \| \bu(t) - \bu_{\texttt{leader}}(t) \|^2  dt, \  \bu = [x(t), y(t)]^T , \  \bp = [v(t), \omega(t)]^T
\end{equation}
subject to dynamical constraints \eqref{eq:aircraft_dyn} and inequality
constraints
\begin{equation}
  v_{\texttt{min}} \leq  v(t) \leq v_{\texttt{max}}, \quad \omega_{\texttt{min}} \leq  \omega(t) \leq \omega_{\texttt{max}}.
\end{equation}

This is a simple example from \cite{Raffard2005} but has all the complexities
including nonlinearity and inequality constraints that are common for practical
dynamical optimal control applications. We implemented this example in
\texttt{PETSc} using \texttt{PETSc} time integrators for solving the dynamical
system and using \texttt{TAO} for optimization. For optimization, we use the
exact Newton method and the classic limited-memory
Broyden--Fletcher--Goldfarb--Shanno (BFGS) method in \texttt{TAO}. The
first-order derivative information (that is, the gradient $\nabla_{\bp} \psi$)
required by both methods is obtained with the first-order adjoint solver, while
the second-order derivative information required by the exact Newton method is
obtained with the second-order adjoint solver and provided in a matrix-free
form (as the Hessian-vector product
$\nabla^2_{\bp} \psi \, \bsigma$). The bound constraints are handled by using an active-set approach
\cite{dener2020tao,Dener2019} in which the problem is reduced to an
unconstrained minimization problem and the descent direction is searched by the
projected line search.

Figure \ref{fig:ex1_conv} shows that the second-order derivative calculated with
the \texttt{PETSc} adjoint solver speeds up the convergence of the optimization
significantly: the exact Newton method takes $7$ iterations to drive the norm of
the gradient of the objective function below $10^{-13}$, whereas the BFGS method
\cite{Benson01alimited} approaches $10^{-8}$ after $50$ iterations.

To validate the gradient computed with the adjoint solver, we leverage the
feature of automatically comparing the gradient with the finite-difference
approximation in \texttt{TAO}, and we perform a analogous test of the Taylor
remainder convergence test in \cite{Farrell2013}. While the comparison itself
can indicate the quality of the adjoint solution, the convergence test
shows the consistency between the forward model and the adjoint
model. By observing that
\begin{subequations}
\begin{align}
   \psi(\bu,\bp+h \tilde{\bp}) - \psi(\bu,\bp) \rightarrow 0 \text{
     with } \mathcal{O} (h)\,,~\textnormal{and}\\
   \psi(\bu,\bp+h \tilde{\bp}) - \psi(\bu,\bp) - h \tilde{\bp} \nabla_{\bp} \psi \rightarrow 0 \text{ with } \mathcal{O} (h^2) ,
\end{align}
\end{subequations}
we find that the difference between the gradient approximated by using central finite-difference and the adjoint solution converges at second order:
\begin{equation}
  \| \nabla_{\bp} \widetilde{\psi} - \nabla_{\bp} \psi  \|  \rightarrow 0 \text{ at } \mathcal{O} (h^2).
\end{equation}

Table \ref{tab:fd_validation} shows the results of the convergence test. As
expected, a second-order convergence is achieved, indicating that the adjoint
solution is correct.

\begin{table}
  \caption{Convergence test for the gradient computed with \texttt{TSAdjoint}.
  It is performed for the first gradient calculation in the optimization loop.
  RK4 is used for time integration.}
  \label{tab:fd_validation}
  \centering
  \resizebox{0.4\textwidth}{!}{
  \begin{tabular}{ c c c }
    \toprule
    h &  $\|\nabla_p \psi - \widetilde{\nabla}_p \psi\|$  &  order \\
    \midrule
    0.005 &  3.415e-6  &  \\
    0.0005 & 3.416e-8  & 2 \\
    0.00005 & 3.334e-10 & 2 \\
  \bottomrule
  \end{tabular}
  }
\end{table}

\subsection{An inverse initial value problem}
\begin{figure}
  \centering
  \begin{subfigure}{0.32\linewidth}
    \includegraphics[width=\textwidth]{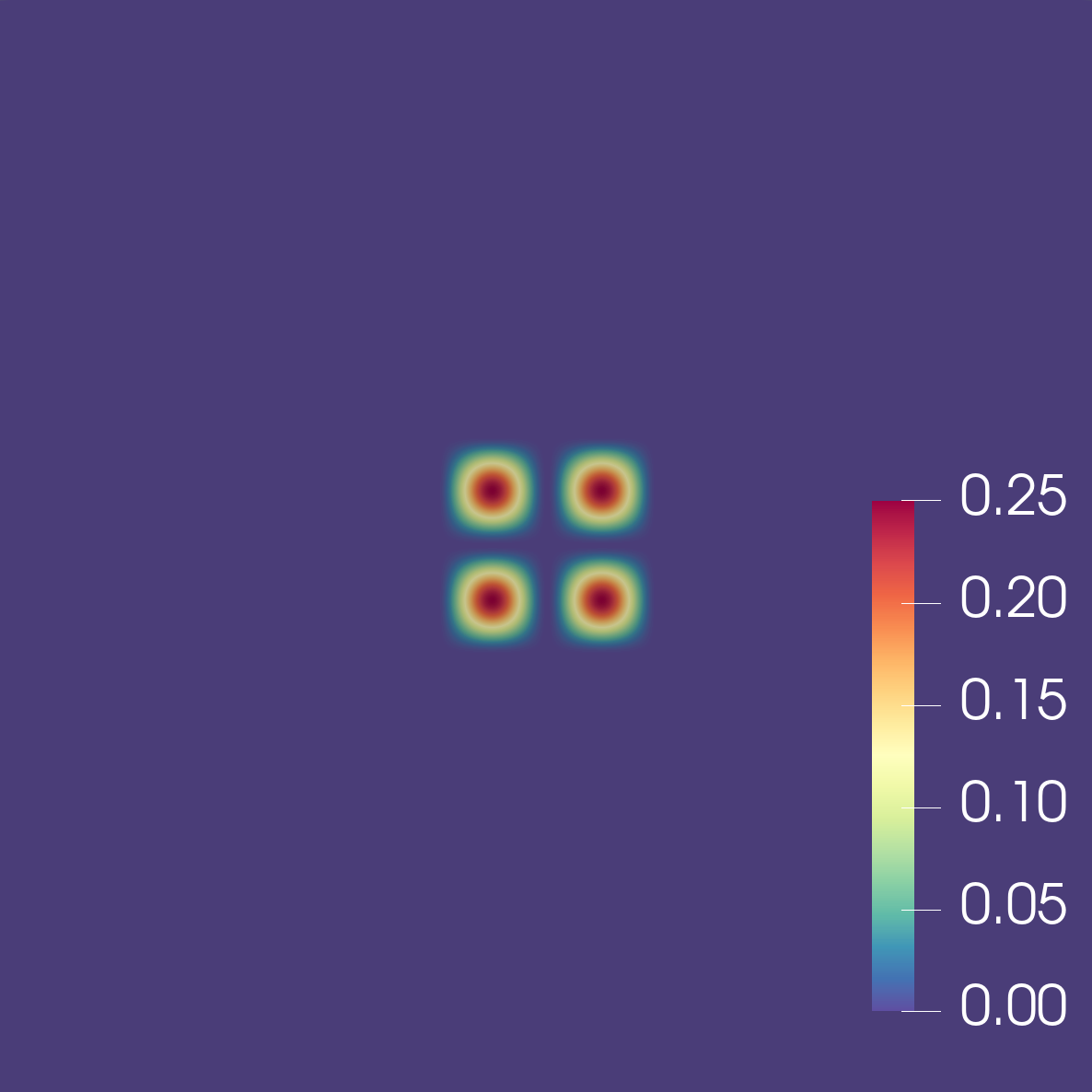}
    \caption{t=0 sec}
  \end{subfigure}
  \begin{subfigure}{0.32\linewidth}
    \includegraphics[width=\textwidth]{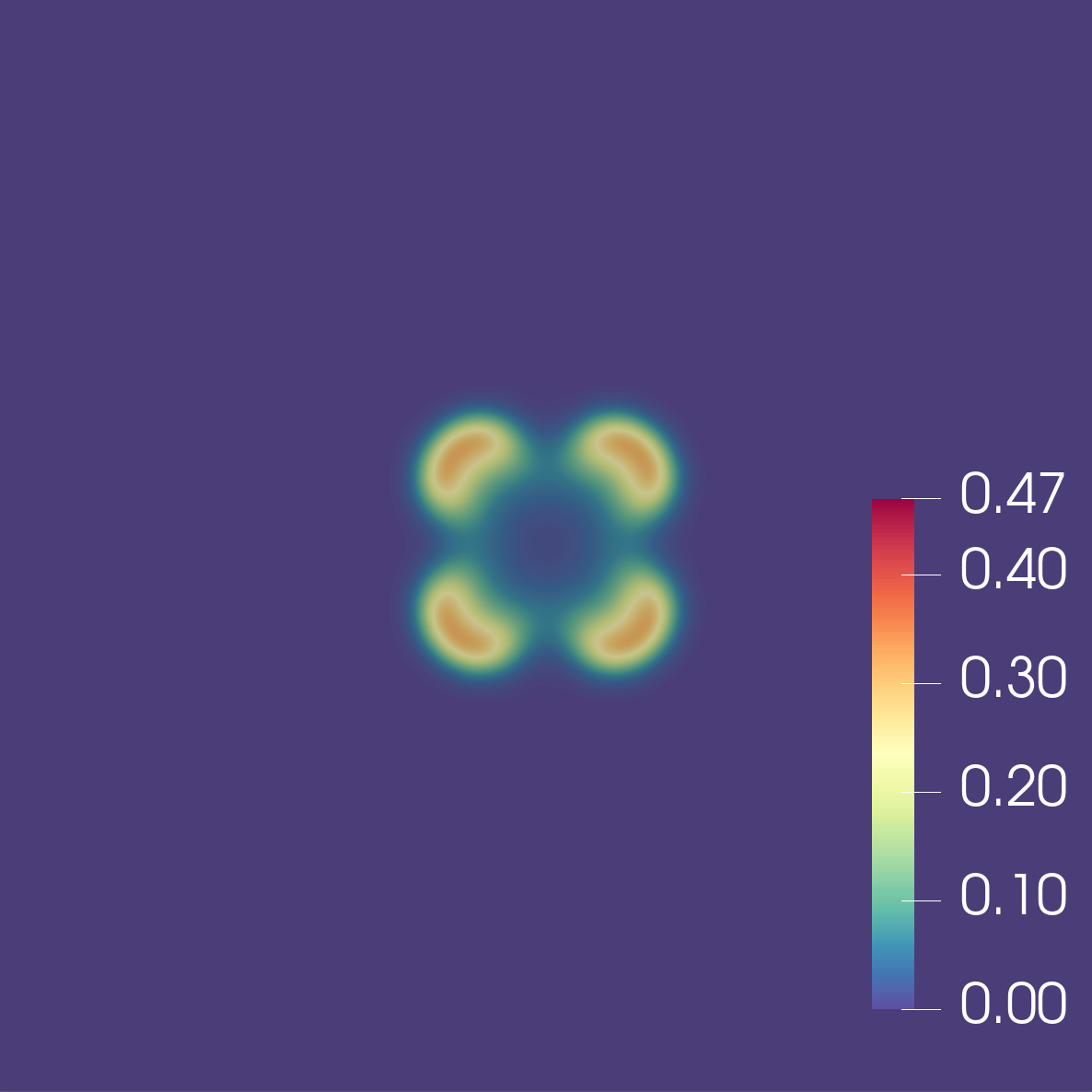}
    \caption{t=100 sec}
  \end{subfigure}
  \begin{subfigure}{0.32\linewidth}
    \includegraphics[width=\textwidth]{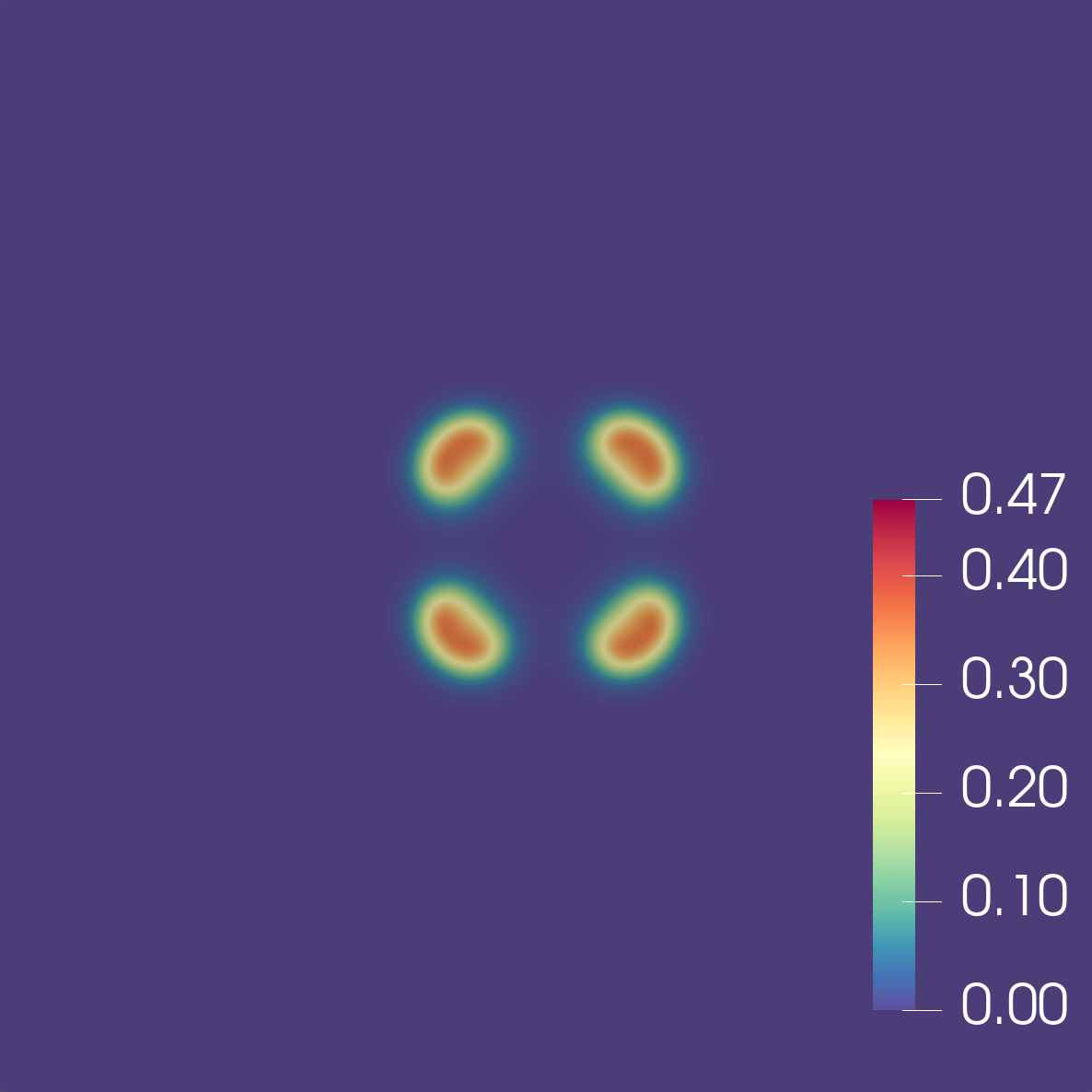}
    \caption{t=200 sec}
  \end{subfigure}
  \caption{Evolving spatial patterns of the concentrations $v$ in the Gray--Scott equations.}
  \label{fig:pattern}
\end{figure}

This example demonstrates the application of adjoint methods in an inverse
problem of recovering the initial condition for a time-dependent PDE and illustrates the
parallel performance of the adjoint calculation involved. The problem can be
formulated as a PDE-constrained optimization problem that minimizes the $L_2$
norm of the discrepancy between simulated and observed results:
\begin{equation}
  \mathop{\text{minimize}}_{\bU_0} \| \bU(t_f) - \bU^{ob}(t_f)\|^2
\end{equation}
subject to the Gray--Scott equations \cite{hundsdorfer2007}
\begin{equation}
  \begin{aligned}
    \dot{\mathbf{u}} = D_1 \nabla^2 \mathbf{u} - \mathbf{u} \mathbf{v}^2 + 
\gamma \, (1 -\mathbf{u}) \\
    \dot{\mathbf{v}} = D_2 \nabla^2 \mathbf{v} + \mathbf{u} \mathbf{v}^2 -
(\gamma + \kappa)\,\mathbf{v}, \\
  \end{aligned}
  \label{eq:diffusionreaction}
\end{equation}
where $\bU = [\mathbf{u} \, \mathbf{v}]^T$ is the PDE solution vector and $\bU_0$ is the initial condition. The PDE models the reaction and diffusion of two interacting species that produce spatial patterns over time, as shown in Figure \ref{fig:pattern}.

% perturbed $\sin^2{(2\pi x}) \cos^2{(2\pi y)}/4$ The adjoint solution
% corresponds to the sensitivities of one component in the final solution with respect to.
% the initial conditions.
In our simulation, the PDE is solved with the method of lines. A
centered finite-difference scheme is used for spatial discretization. The
computational domain is $\Omega \subset [0,2]^2$. The time interval is $[t_0,t_f] =
[0,5]$. A reference solution is generated from the initial condition
\begin{equation}
  \mathbf{u}_0 = 1 - 2 \mathbf{v}_0, \quad
  \mathbf{v}_0  = 
  \begin{cases}
    \sin^2{(4\pi x}) \cos^2{(4\pi y)}/4,\; \forall x,y \in [1.0,1.5] \\
    0 \; \text{otherwise}
    \end{cases}
\end{equation}
and set as observed data. The nonlinear system that arises at each time
step is solved by using a Newton-based method with line search. 
%By default the
%linear systems are solved using GMRES \cite{Saad1986} with the block Jacobi
%preconditioner (and ILU(0) for each block).
For large-scale experiments, we use
the geometric algebraic multigrid (GAMG) preconditioner with the following
options:
\begin{lstlisting}
-mg_levels_ksp_type richardson -mg_levels_pc_type jacobi
-pc_gamg_process_eq_limit 500 -pc_gamg_square_graph 10
-pc_gamg_reuse_interpolation -pc_gamg_repartition false
\end{lstlisting}
The last two options allow us to reuse the prolongation operator and avoid
repartitioning across the iterations, thus mitigating the performance impact of
the setup phase, which is complex and difficult to scale for algebraic multigrid
preconditioners \cite{Baker2011}.

To solve the optimization problem, we use the limited-memory BFGS algorithm
\cite{Benson01alimited} in \texttt{TAO} \cite{Dener2019} by providing to
\texttt{TAO} a function that returns the value of the objective function and its
gradient with respect to $U_0$ in the Euclidean space (thus 
implying mesh dependence
\cite{Schwedes2017}). The function is computed with a forward solve solving the
PDE for solution and evaluation of the objective function, followed by an
adjoint solve calculating the gradient expressed by
\eqref{eq:tot:phi:optimization}. The gradient is verified by using an analog of
the Taylor convergence test described in Section \ref{sec:optimal_control_ex}.
As shown in Table \ref{tab:fd_validation_pde}, the theoretical order of
convergence $2$ is achieved, indicating that the adjoint solution is correct.

\begin{table}
  \caption{Convergence test for the gradient computed with \texttt{TSAdjoint}.
  It is performed for the first gradient calculation in the optimization loop.
  The grid size is set to $50 \times 50$. RK4 is used for time integration.}
  \label{tab:fd_validation_pde}
  \centering
  \resizebox{0.5\textwidth}{!}{
  \begin{tabular}{ c c c }
    \toprule
    h &  $\|\nabla_{\bU_0} \psi - \nabla_{\bU_0} \widetilde{\psi}\|$  &  order \\
    \midrule
    0.005 &  2.649e-3   &  \\
    0.0005 & 2.648e-5 & 2 \\
    0.00005 & 2.649e-7 & 2 \\
  \bottomrule
  \end{tabular}
  }
\end{table}

\textbf{Efficiency} The efficiency of the adjoint solver can be defined by the
ratio of the cost of the forward solve to the cost of the adjoint solve. The
results for three timestepping methods are presented in Table
\ref{tab:diffusiont_reaction}. 
\begin{table}
  \caption{Performance comparison of two different Jacobian evaluation strategies and three selected timestepping methods. The grid size used in the tests is $100 \times 100$. A fixed stepsize of $0.5$ is used on the time interval $[0,5]$.}
  \label{tab:diffusiont_reaction}
  \centering
  \resizebox{\textwidth}{!}{
  \begin{tabular}{ c c c c c c c c}
    \toprule
    \multirowcell{2}{Jacobian} &  \multirowcell{2}{Time\\integration}  & \multirowcell{2}{Wall time\\ (second)} & \multirowcell{2}{Ratio\\(adjoint/forward)} & \multirowcell{2}{Iterations} & \multirowcell{2}{First-order \\ computations} & \multirowcell{2}{RHS\\evaluations} & \multirowcell{2}{Jacobian\\evaluations} \\
    & & & & & & \\
    \midrule
    \multirowcell{3}{Analytical} & Backward Euler & 30.0 & 0.48 &188 & 194 & 5,870 & 5,870 \\
     & Crank--Nicolson  & 45.4 & 0.76 & 253 & 264 & 10,581 & 10,581 \\
     & Runge--Kutta 4 & 25.6 & 38.03 & 246 & 253 & 10,120 & 10,120 \\
     \midrule
    \multirowcell{3}{FDColoring} & Backward Euler & 19.9 & 0.48 & 188 & 196 & 67,190 & - \\ 
    & Crank--Nicolson & 28.8 & 0.66 & 246 & 254 & 127,252 & - \\
    & Runge--Kutta 4 & 11.8 & 16.48 & 244 & 255 & 122,400 & - \\
    \midrule
    \multirowcell{3}{Matrix-free} & Backward Euler & 4.3 & 0.40 & 186 & 194 & 5,869 & - \\ 
    & Crank--Nicolson & 5.1 & 0.41 & 240 & 246 & 9,865 & - \\
    & Runge--Kutta 4 & 1.8 & 1.11 & 229 & 237 & 9,480 & - \\
  \bottomrule
  \end{tabular}
  }
\end{table}

The two selected implicit methods, backward Euler and Crank--Nicolson, are
special cases of the theta method ($\theta = 1/2$ for backward Euler and $
\theta = 0$ for Crank--Nicolson). Both achieve an efficiency ratio of less than
$1$. For linear problems, the optimal ratio is $1$, assuming the cost of
assembling the linear system and the right-hand side is identical and the cost
of solving the transposed linear system in an adjoint time step is equivalent to
the cost of solving the system in the corresponding forward time step. For
nonlinear problems, a smaller ratio is expected because the forward solve
requires the solution of one or more (depending on the timestepping algorithm)
nonlinear systems while the adjoint run requires only the solution of linear
systems at each adjoint time step, the number of which is the same as the number
of nonlinear systems required in the forward time step. In this example, the
nonlinear solve takes $2$ Newton iterations on average. The adjoint solver based
on the backward Euler scheme is slightly more efficient than the adjoint solver
based on the Crank--Nicolson scheme because the Jacobian evaluation needed in
equation \eqref{eqn:disadj_theta_simple_b} can be avoided for backward Euler
when the mass matrix is the identity. This kind of performance optimization can
be discovered easily from the formula and implemented; however, it is difficult
to be realized by algorithmic differentiation tools.

The fourth-order explicit method, Runge--Kutta 4, has a relatively high
efficiency ratio when using the explicit Jacobian, mainly because the right-hand
side evaluation is significantly faster than the Jacobian evaluation, which
consists of costly memory operations including assembling the matrix. 
When using the matrix-free approach, however, the efficiency ratio is significantly
improved for Runge--Kutta 4, while the ratios for the other two methods tested
within \texttt{PETSc} are slightly improved.

% Note that the Jacobian does not have to be provided explicitly, \texttt{PETSc}
% supports the matrix-free Jacobian for which users only need to implement the
% application of the Jacobian to a given vector. Despite the high efficiency
% ratio, the explicit method is still the most efficient option for this problem
% since the (transposed) matrix-vector multiplication needed in the adjoint is
% considerably cheaper than solving an implicit system.

Interestingly, using finite differences and coloring outperforms the analytical
Jacobian for this example and implementation. As Table
\ref{tab:diffusiont_reaction} indicates, the number of iterations of the
optimization process does not vary much between the two choices. The Jacobian
approximation takes $10$ right-hand side function evaluations ($5$ colors and 2
components in the PDE). Although finite differences need more arithmetic operations, the array of values generated from the approximation can be
transferred into a \texttt{PETSc} sparse matrix efficiently. In contrast, in the
implementation of the analytical Jacobian, the matrix values are set row-wise,
which is natural for sparse matrices in the compressed sparse row format but
less cache-efficient.

\begin{figure}
  \centering
  \begin{subfigure}{0.48\linewidth}
    \includegraphics[width=\textwidth]{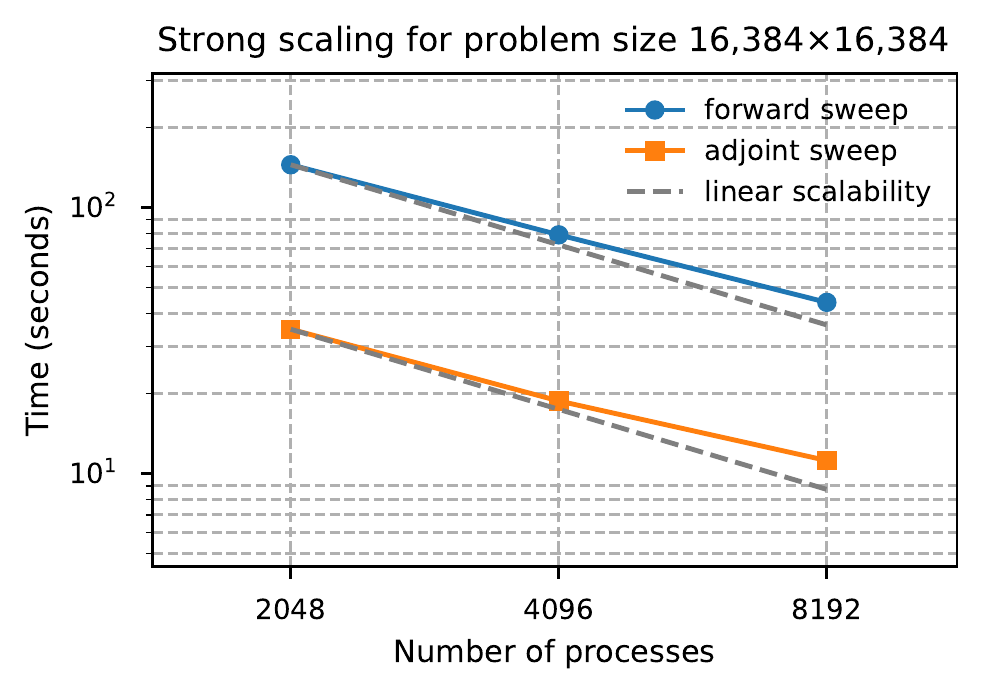}
    \caption{Backward Euler}
  \end{subfigure}
  \begin{subfigure}{0.48\linewidth}
    \includegraphics[width=\textwidth]{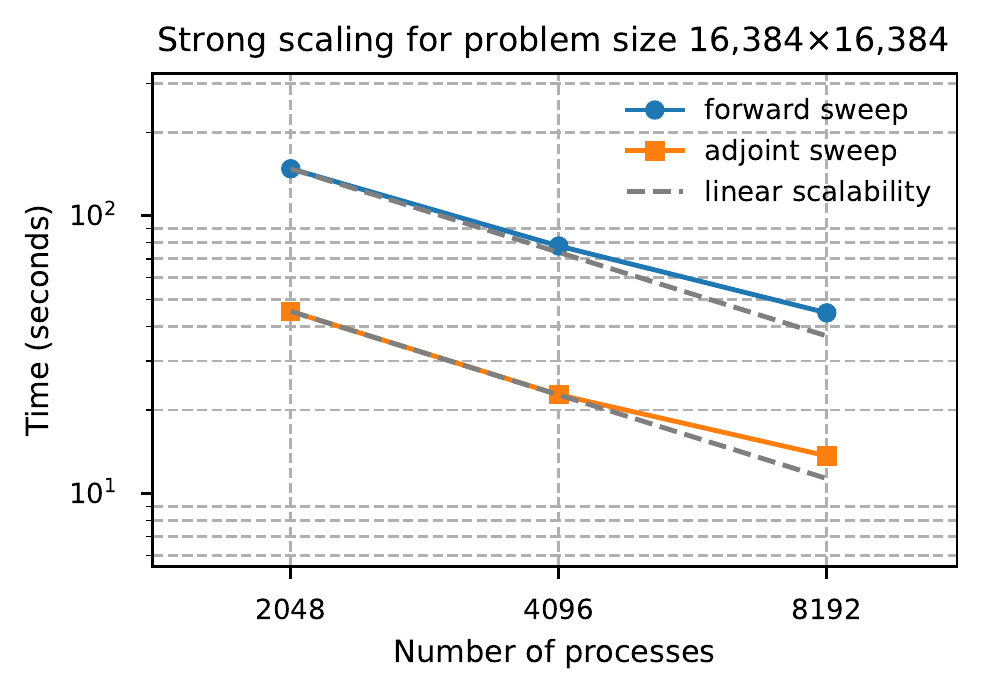}
    \caption{Crank--Nicolson}
  \end{subfigure}
  \begin{subfigure}{0.48\linewidth}
    \includegraphics[width=\textwidth]{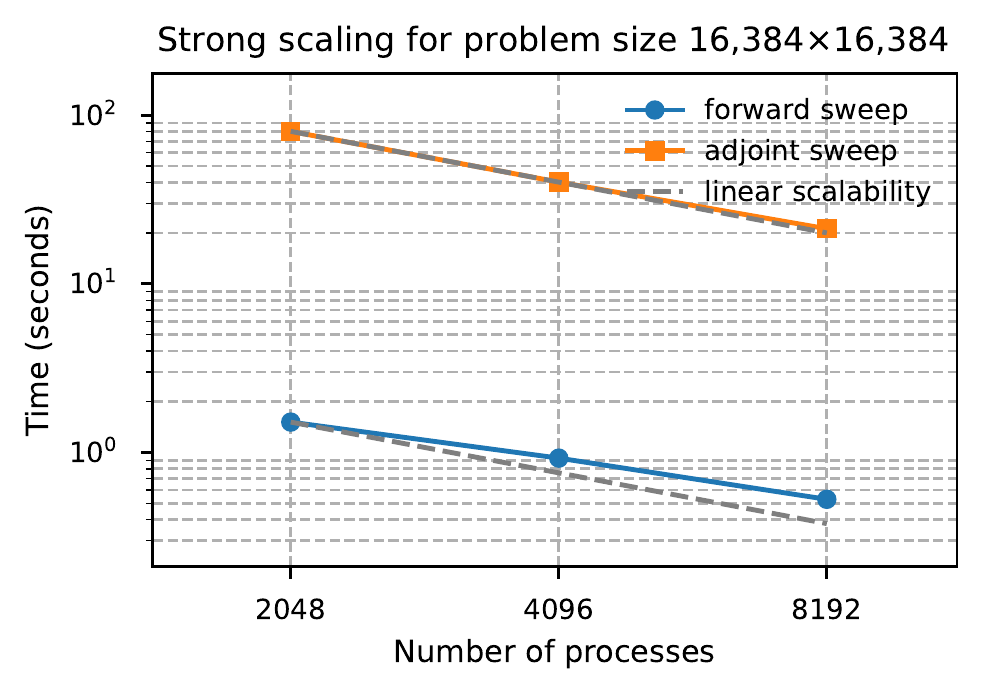}
    \caption{Runge--Kutta 4}
  \end{subfigure}
  \caption{Strong scaling of the adjoint sensitivity calculation for the 2D reaction-diffusion equation \eqref{eq:diffusionreaction} on NERSC's supercomputer Cori. In all the tests, $32$, $64$, and $128$ compute nodes with $64$ MPI processes on each node are used. The grid size is $16,384 \times 16,384$ (yielding about $0.5$ billion degrees of freedom). Three time integrations methods are tested.}
  \label{fig:strongscaling_ex5adj}
\end{figure}

%the Galerkin coarse grid operator construction ==RAP
\textbf{Parallel scaling.} To demonstrate the scalability of the adjoint solver,
we ran the gradient calculation portion of this benchmark problem with fine grid
resolution on the Intel Xeon Phi Knights Landing (KNL) nodes of the NERSC
supercomputer Cori. Each KNL node is assigned $64$ MPI processes with one
process per core. Manually optimized linear algebra kernels (e.g., vectorized
matrix-vector multiplication \cite{Zhang2018}) are used for the best
performance. Figure \ref{fig:strongscaling_ex5adj} shows the scaling results for
up to $8,192$ MPI processes. Backward Euler and Crank--Nicolson exhibit good
parallel scaling for both the forward solve and the adjoint solve. In fact, the
solve phase of GAMG scales well in the strong-scaling regime. The setup phase
scales up to $40,96$ processes. For Runge--Kutta 4, the scaling of the forward
solve is not ideal because the communication cost in the right-hand side
function becomes relatively large compared with the computation time as the
number of processes increases. For the adjoint solve, however, perfect linear
scaling is observed.

\subsection{A Firedrake example: adjoint of Burgers' equation}
\begin{figure}
  \centering
  \begin{subfigure}{0.32\linewidth}
    \includegraphics[width=\textwidth]{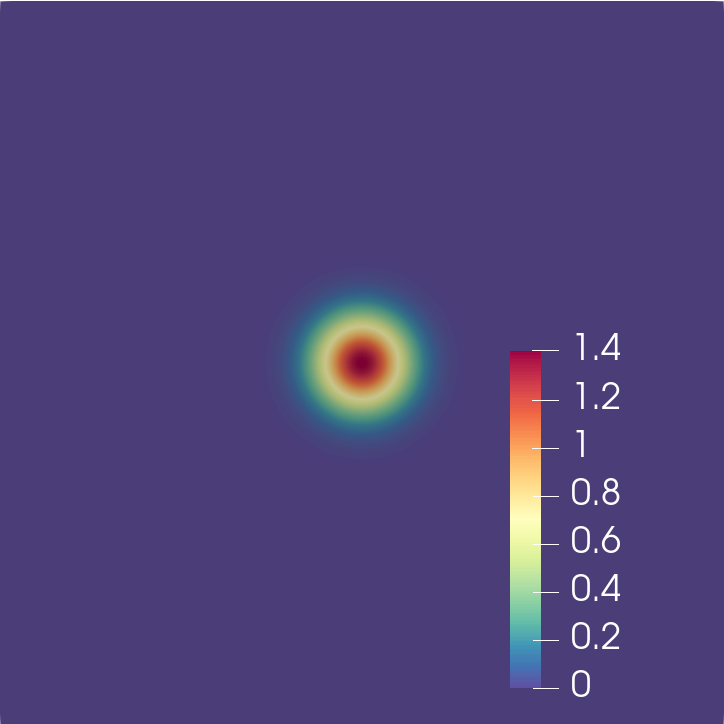}
    \caption{}
  \end{subfigure}
  \begin{subfigure}{0.32\linewidth}
    \includegraphics[width=\textwidth]{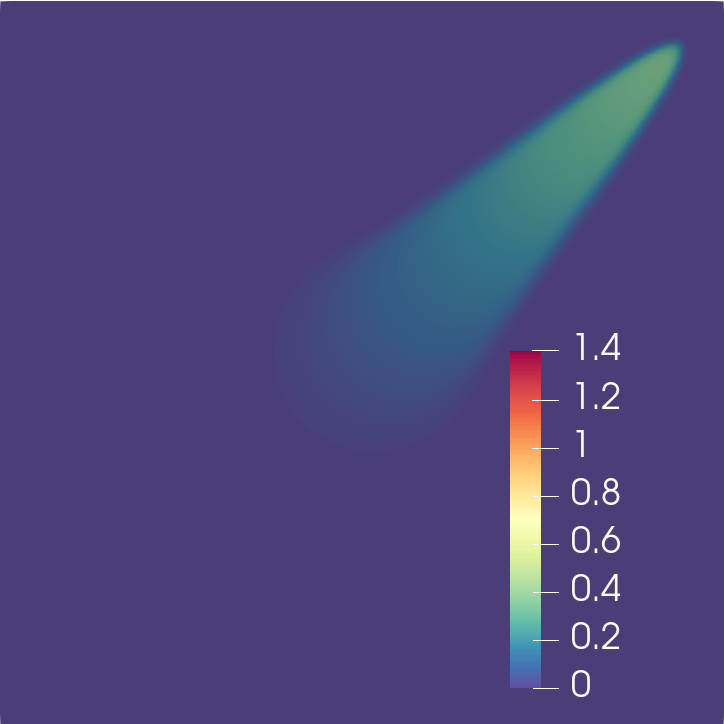}
    \caption{}
  \end{subfigure}
  \begin{subfigure}{0.32\linewidth}
    \includegraphics[width=\textwidth]{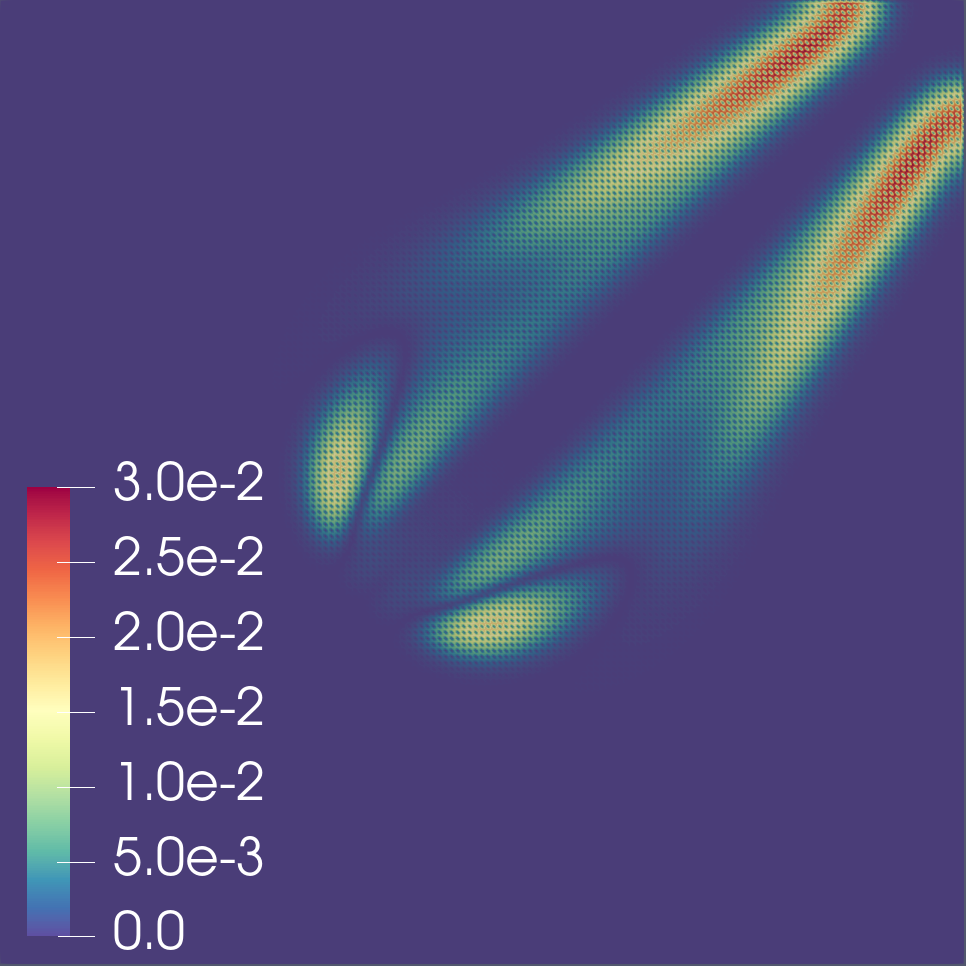}
    \caption{}
  \end{subfigure}
  \caption{Initial condition (a), final solution at $T=2s$ (b), and sensitivity of the $H^1$ error norm with respect to the initial condition in the Lagrangian function space (c).}
  \label{fig:burgers}
\end{figure}

We consider Burgers' equation on a uniform square mesh:
\begin{equation}
  \begin{aligned}
  \dot{\bU} + (\bU \cdot \nabla) \bU - \nu \Delta \bU = 0 \\
  (n\cdot \nabla) \bU = 0 \ \textrm{on}\ \Omega ,
  \end{aligned}
\end{equation}
where $\Omega$ is the domain and $\nu$ is a constant scalar viscosity. The
equation is discretized in space by using Lagrange finite elements of polynomial
degree 2. The initial condition is a Gaussian profile with amplitude $1.0$ and
distribution width $0.06$,  as shown in Figure \ref{fig:burgers}; and $16$ uniform
time steps are used on the time interval $[0,2]$ seconds. For testing, we
compute the sensitivity of the $H^1$ error norm of the solution in the
finite-element function space with respect to the initial condition:
\begin{equation}
  \int_{\Omega} \left( (\bU-\bUref) \cdot (\bU-\bUref) + (\nabla \bU- \nabla \bUref) \cdot (\nabla \bU- \nabla \bUref) \right)  \mathrm{d} x ,
\end{equation}
where the reference solution $\bUref$ is computed by using a strict stepsize and a
fine mesh. This example is implemented by using only a few lines of Python code.
The right-hand side function and Jacobian function defining the ODE problem are
automatically generated by specifying the variational formulations of the
semi-discretized PDE using Firedrake; they are provided to the \texttt{PETSc}
timestepping solver through \texttt{petsc4py} \cite{Lisandro2011} for the
forward and the adjoint solution.

Table \ref{tab:burgers_performance} lists the total runtime and number of
right-hand side and Jacobian evaluations for both the forward and the adjoint
computation. We observe that the adjoint-to-forward ratios are $0.25$ for
backward Euler and $0.14$ for Crank--Nicolson. While the runtime of the forward
solve differs significantly for the time integration methods, the runtime of the
adjoint solve is approximately the same. The reason is that the right-hand side
function evaluation (the spatial discretization) dominates the total
computational cost, while the adjoint solver of backward Euler or Crank--Nicolson
requires the same number of Jacobian evaluations and the same number of linear
solves (one per adjoint time step).

\begin{table}
  \caption{Timings of  Burgers' adjoint in Firedrake.}
  \centering
  \resizebox{0.99\textwidth}{!}{
  \begin{tabular}{ c c c c c c}
    \toprule
    \multirowcell{2}{Time integration}  & \multirowcell{2}{Stage} &  \multirowcell{2}{Wall time \\(second)} & \multirowcell{2}{Ratio} & \multirowcell{2}{RHS\\evaluations} & \multirowcell{2}{Jacobian\\evaluations} \\
    & & & & & \\
    \midrule
    \multirowcell{2}{Backward Euler} &  forward run & 22.386 & 1 & 142 & 109 \\
                        & reverse run & 5.543 & 0.25 & 0 & 32\\
    \midrule
    \multirowcell{2}{Crank--Nicolson} &  forward run & 40.704 & 1 & 254 & 157\\
                        & reverse run & 5.868  & 0.14 & 0 & 32\\
  \bottomrule
  \end{tabular}
  }
  \label{tab:burgers_performance}
\end{table}

\section{Conclusion}
Algorithmic differentiation has long been needed by many scientific
applications, especially as machine learning becomes increasingly popular. It
has been realized at different abstraction levels, posing different
challenges for application developers and software developers. The new tool
presented in this paper, \texttt{PETSc} \texttt{TSAdjoint}, provides an
efficient and accurate approach for computing first-order and second-order
adjoints for ODEs, DAEs, and time-dependent nonlinear PDEs. It makes the task of
gradient calculation easier by avoiding full differentiation of the entire code,
with no loss of accuracy and speed. Minimal changes are required for
applications using \texttt{PETSc} time integrators to be equipped with
sensitivity analysis capabilities. An optimal checkpointing component has been
developed to deliver transparent and optimal checkpointing strategies on
high-performance computing platforms. Parallelism is inherited from
\texttt{PETSc} parallel infrastructures. Thanks to the hierarchical structure of
\texttt{PETSc}, the adjoint solvers take advantage of the well-developed
nonlinear and linear iterative solvers and the extensive collection of
preconditioners in \texttt{PETSc}. 

Extensive experiments have been performed to demonstrate the usability,
efficiency, and scalability of the adjoint solvers. We have shown that they can
be easily used with various other scientific computing libraries or tools in
different programming languages. We have also shown that using finite
differences and coloring and relying on high-level AD are efficient and
convenient alternatives to deriving and implementing an analytical Jacobian. For
first-order adjoints, the adjoint solve cost is typically less than the forward
solve cost when implicit timestepping methods are employed. The performance ratio for
explicit methods can exceed 1 if the Jacobian matrix is provided in the explicit
form; however, this could be mitigated by using matrix-free implementations.
Furthermore, the adjoint computation of PDEs scales nicely to large numbers of
cores on a supercomputer, even when the scaling of the forward solve is not
ideal. In addition, we show how the second-order adjoint sensitivities can be
used to accelerate the convergence of optimization in an optimal control
problem. Without a doubt, Hessian-related information for  the dynamical system
is needed by second-order adjoints and may be difficult to compute. However,
\texttt{PETSc} \texttt{TSAdjoint} requires only a rank-1 vector-Hessian-vector
product for the second-order adjoints.

As far as we know, this library is the first general-purpose HPC-friendly
library that offers first-order and second-order discrete adjoint capabilities
based on multistage time integration methods, supports sensitivity analysis for
hybrid dynamical systems, and comes with sophisticated checkpointing support
that is transparent to users. We expect that more applications in
PDE-constrained optimization, data assimilation, uncertainty quantification, and
machine learning will be enabled by our development.

\section*{Acknowledgments}
We are grateful to Jed Brown for his extensive code review and helpful
discussions during the software development. We also thank Mark Adams for the
guidance on GAMG and Todd Munson for his valuable comments on the draft.

\bibliography{tsadjoint}
\bibliographystyle{siamplain}

% Argonne Licence.
\iffalse
\vfill
\begin{flushright}
\framebox{\parbox{0.9\textwidth}{The submitted manuscript has been created by
UChicago Argonne, LLC, Operator of Argonne National Laboratory (``Argonne'').
Argonne, a U.S. Department of Energy Office of Science laboratory, is operated
under Contract No. DE-AC02-06CH11357. The U.S. Government retains for itself,
and others acting on its behalf, a paid-up nonexclusive, irrevocable, worldwide
license in said article to reproduce, prepare derivative works, distribute
copies to the public, and perform publicly and display publicly, by or on behalf
of the Government. The Department of Energy will provide public access to these
results of federally sponsored research in accordance with the DOE Public Access
Plan. http://energy.gov/downloads/doe-public-access-plan. }}
\normalsize
\end{flushright}
\fi
\end{document}